\documentclass[journal]{IEEEtran}
\usepackage[T1]{fontenc}
\ifCLASSINFOpdf
   \usepackage[pdftex]{graphicx}
   \DeclareGraphicsExtensions{.pdf,.jpeg,.png}
\else
\fi
\usepackage{cite}
\usepackage{graphicx}
\usepackage[cmex10]{amsmath}
\usepackage{algorithm}
\usepackage{algorithmic}
\usepackage{bm}
\usepackage{amssymb}
\usepackage{breqn}
\usepackage{array}
\usepackage[caption=false,font=normalsize,labelfont=sf,textfont=sf]{subfig}
\usepackage{url}
\usepackage{placeins}
\usepackage{multirow}
\usepackage{lipsum}
\usepackage{mathtools, cuted}
\usepackage{multicol, blindtext}
\usepackage{amsthm}

\newtheorem{theorem}{Theorem}
\usepackage{threeparttable}
\newtheorem{Lemma}{Lemma}
\begin{document}
\title{Beamforming and Resource Allocation for Multi-User Full-Duplex  Wireless Powered Communications in IoT Networks}
\author{\IEEEauthorblockN{Derek~Kwaku~Pobi~Asiedu,~\IEEEmembership{Member,~IEEE}, Sumaila Mahama,~\IEEEmembership{Student Member,~IEEE}, Changick Song,~\IEEEmembership{Senior Member,~IEEE}, Dongwan Kim,~\IEEEmembership{Member,~IEEE}, and Kyoung-Jae~Lee,~\IEEEmembership{Member,~IEEE}}
\thanks{This work was supported in part by the Institute for Information and Communications Technology Promotion (IITP) grant through Korea Government (MSIT) (2018-0-00812, IoT Wireless Powered Cognitive Radio Communications With User-Centric Distributed Massive MIMO Systems), and in part by Korea Evaluation Institute of Industrial Technology (KEIT) grant through the Korea government (MOTIE) (20001056, Development of Low-Power Massive MISO SWIPT for IoT Wearable). The material in this paper was presented in part at the IEEE VTC, Porto, Portugal, 2018\cite{Asiedu18}. (\textit{Corresponding author: Kyoung-Jae Lee.})} 
\IEEEcompsocitemizethanks{\IEEEcompsocthanksitem Derek~Kwaku~Pobi~Asiedu and Kyoung-Jae~Lee are with the Department of Electronics and Control Engineering, Hanbat National University, Daejeon 34158, South Korea. (email: kyoungjae@hanbat.ac.kr)\protect

S. Mahama is with the Department of Electronics Engineering, University of York, Heslington York YO10 5DD, United Kingdom.\protect

C. Song is with the Department of Electronics Engineering, Korea National University of Transport, Chungju 27469, South Korea.\protect

D. Kim is with the Department of Electronics Engineering, Dong-A University,  Busan 49315, South Korea.\protect
}
}

\maketitle
\begin{abstract}
For a self-sustaining wireless communication system in Internet-of-Things (IoT) networks, energy harvesting (EH) can be implemented at each user node as a constant renewable power supply source. Hence, an investigation into the use of wireless powered communication network (WPCN) protocols to facilitate communication between an access point (AP) and multiple mobile users (MUs) is presented in this paper. The AP has multiple antennas and operates in the full-duplex (FD) mode. The MUs, on the other hand, have single antennas and works in the half-duplex (HD) mode. Each MU communicating with the FD-AP is assigned to one of two groups, based on the time allocation and channel access for either uplink (UL) or downlink (DL) communication. The channel assignment, time resource, and power resource allocations are optimized to maximize the UL weighted sum-rate. The sum-rate optimization problem is found to be non-convex. Therefore, an iterative algorithm is investigated to optimize the UL weighted sum-rate of the propose FD-WPCN system. Next, the proposed FD-WPCN algorithm is modified for a HD-WPCN enabled communication between the AP and multiple MUs. Extensive simulations are conducted to verify the proposed algorithm for FD-WPCN and compare its performance with the HD-WPCN counterpart. From the simulation results, FD-WPCN outperformed HD-WPCN at a low AP transmit signal-to-noise ratio (SNR) region. The opposite behavior is observed for high AP transmit SNR due to increasing residual self-interference at the FD-AP.
\end{abstract}
\IEEEpeerreviewmaketitle
\section{Introduction}
\label{secintro}
\IEEEPARstart{E}XPECTATIONS for next-generation wireless communication systems consist of having a better quality-of-service (QoS) in terms of high data rates and a self-sustaining wireless communication ecosystem \cite{Chen15,Huang015,Jang15,Huang15}. A full-duplex (FD) cellular system promises to potentially increase system capacity (i.e., improved data rates) compared to half-duplex (HD) cellular systems for next-generation cellular systems \cite{Song15,Sabh14,Anokye18}. While HD systems receive and transmit wireless signals over different frequency bands, FD systems transmit and receive signals over the same frequency band \cite{Song15,Sabh14,Anokye18}. Two different antenna structures can be used to achieve FD communications. The first is the shared antenna configuration, in which a single antenna simultaneously receives and transmits signals via a three-port system circulator. In the second configuration, known as the separated antenna structure, information is transmitted and received over different antennas \cite{Anokye18}. Even though FD systems have the potential of doubling the achievable rate of a communication system, one major drawback with FD systems is the existence of self-interference (SI) \cite{Song15,Sabh14,Anokye18}. SI is the interference that the FD system node's transmitter causes to its receiver due to the simultaneous transmission and reception of information signals \cite{Song15,Sabh14,Anokye18}. The SI can be mitigated by antenna design, and digital and analog cancellation techniques \cite{Jang15,Huang15,Song15,Sabh14,Anokye18}. 

Next-generation wireless communication systems are expected to support Internet-of-Things (IoT) networks, in which a myriad of network-enabled devices interact with each other over the wireless broadcast channels and internet gateways \cite{Da14,Oteafy17,Deng18,Adhatarao18,Raghavendra06}. In IoT networks, it is essential to build a self-sustaining system to reduce maintenance costs \cite{Da14,Oteafy17,Deng18,Adhatarao18}. In particular, self-powered devices that can harvest energy from an external energy source will significantly reduce the need to charge the batteries of wireless IoT devices \cite{Yick08,Ciuonzo12,Ciuonzo14,Rossi16,Yetgin17}. Solar, hydro, wind, vibrations are traditional electrical energy sources, but their performance depends on natural constraints such as weather, and thus are hard to control \cite{Chen15,Huang015,Mahama17,Asiedu018}. Recently, the radio frequency (RF) signal energy is gathering great attention as a more controllable electrical energy source. In this case, energy beacon such as the base station or access point can intentionally transfer power to the mobile users through the RF signals. This is known as wireless power transfer (WPT) \cite{Ulukus15,Ding15,Guo16,Chen15,Huang015,Mahama17,Asiedu018}. 

WPT can be realized in wireless communication systems by two main techniques: wireless powered communication network (WPCN) and simultaneous wireless information and power transfer (SWIPT) \cite{Chen15,Huang015,Mahama17,Asiedu018}. To facilitate both wireless information transfer (WIT) and WPT in cellular systems concurrently, SWIPT is applied to cellular systems \cite{Rui13,Zhou12,Varshney08}. SWIPT can be implemented using two techniques, the time switching (TS) ratio and the power splitting (PS) ratio \cite{Chen15,Huang015,Mahama17,Asiedu018}. On the other hand, in WPCN WPT and WIT occur successively \cite{Chen15,Huang015,Mahama17,Asiedu018}. WPCN is a promising solution to providing stable energy for low-power energy-constrained sensor nodes (i.e., for batteryless sensor nodes or sensor nodes with rechargeable batteries). One key advantage of the WPCN technique is simplicity in WPT design, resulting in a low implementation cost for sensor nodes compared to SWIPT. WPCN also aids in the extension of the IoT network lifespan.

\subsection{Related works}
The authors in \cite{Ju14} have proposed a FD based WPCN system. In \cite{Ju14}, a two-antenna FD hybrid access point (H-AP) communicates with single antenna HD multi-user mobile users (MUs) operating in time-division-multiple-access (TDMA) mode. The H-AP uses one antenna for energy transmission, while the other antenna is used to receive information signals from the MUs. In addition to the two antenna structure, the authors also considered the presence of SI at the H-AP. Each MU in the system model harvests energy when it is not transmitting information signals to the H-AP. A weighted uplink (UL) sum-rate maximization problem was solved based on the time allocations for H-AP downlink (DL) WET and MUs UL WIT, and the H-AP was assumed to have either perfect and imperfect SI cancellation (SIC). The authors also investigated the HD-WPCN case as a baseline for comparison with the FD-WPCN. An iterative algorithm was proposed to determine the time resource allocation that maximizes the UL sum-rate for both the FD-WPCN and HD-WPCN system models. In the simulation section, the FD-WPCN is compared with the HD-WPCN. The simulation showed that the FD-WPCN system architecture outperformed the HD-WPCN system model. Also, the rate trade-offs for two users in both FD-WPCN and HD-WPCN is presented in \cite{Ju14}.

A similar system model to the one in \cite{Ju14} is studied in \cite{Hoon016}, which considers extensive research work on resource allocation for FD-WPCN. With a similar system model and node structures, the difference between the research works presented in \cite{Ju14} and \cite{Hoon016} lies with the consideration of the practical causal energy system in \cite{Hoon016}. Also, the work presented in \cite{Hoon016} covered both a finite and an infinite battery scenario at each MU node. The authors proposed a joint energy and time resource allocation to maximize UL sum-rate. It was assumed that the H-AP is equipped with two antennas (i.e., one antenna acts as an energy transmitter and the other acts as an information receiver), with each MU being fitted with a single antenna. The MUs communicate with the H-AP in a TDMA mode, and the H-AP has perfect SIC. The main focus of the research in \cite{Hoon016} was on the maximization of the system UL sum-rate by optimizing the time resource allocation, and the DL and UL transmit power allocations. In \cite{Hoon016}, two algorithms for the finite and infinite battery cases to maximize the achievable system UL sum-rate were studied and compared. From the simulation results, the infinite battery case outperformed the limited battery case. 

Multi-user device-to-device (D2D) energy minimization for a FD-AP communicating with multiple MUs is investigated in \cite{Yu19}. The work presented in \cite{Yu19}, also has a similar system model as the two previously discussed papers, that is, \cite{Ju14} and \cite{Hoon016}. However, \cite{Yu19} focuses on the H-AP source power minimization based on a given system QoS in terms of required throughput. This differs from the UL sum-rate maximization problems considered in \cite{Ju14} and \cite{Hoon016}. Besides, unlike the work in \cite{Ju14}, perfect SIC is assumed in \cite{Yu19}. However, similar to \cite{Ju14}, \cite{Yu19} considered both the FD-WPCN and HD-WPCN system communication models and TDMA communication mode. From the optimization techniques employed in \cite{Yu19}, two algorithms were proposed, one for the FD-WPCN system model and the other for the HD-WPCN system model. Based on the energy minimization simulation results in \cite{Yu19}, the FD-WPCN required less energy compared to the HD-WPCN system model. Further review and readings on FD-WPCN can be found in \cite{Kang15,Hu17,Lyu19}.

For existing non-energy harvesting (non-EH) wireless multi-user communication systems, practically all users do not communicate with an AP synchronously for their UL and DL communication in a FD communication system. Therefore, while some users are operating in UL, others transmit in DL \cite{Duy16}. For the conventional non-EH wireless FD communication system, \cite{Duy16} proposed a weighted sum-rate maximization scheme by considering both UL and DL communication, channel assignment, and a multiple-input-multiple-out (MIMO) processing capability at the base station (BS) for two groups of users. The users were grouped into two based on their channel assignment and current communication status (i.e., while one group performed DL on a channel, another group performs UL on that same channel, and the reverse occurs on the next channel). An iterative algorithm that maximizes the UL and DL sum-rate for all users based on the UL and DL power allocation, and the channel assignment is proposed in \cite{Duy16}. The proposed FD communication mode is compared to the HD communication mode in the simulation section of the paper. From the simulation results, the FD communication system outperformed the HD mode of communication.

\subsection{Contribution}
This paper studies a new WPCN scenario, where a multi-antenna H-AP serves single antenna MUs in both WIT UL and WPT DL. For the system model covered in this paper, the AP operates in a FD mode, whereas the MUs communicate with the H-AP in a HD mode. The system also operates in a time division duplexing (TDD) mode for the WIT UL and WPT DL communication between the H-AP and MUs. The MUs are in two groups based on whether a MU is communicating with the H-AP in a WIT UL or EH DL mode in a particular time slot. In this paper, the H-AP is assumed to have an imperfect channel-state-information (CSI). The channel assignment, and the time and power resources allocations are jointly optimized to maximize the UL sum-rate. The sum-rate maximization problem is shown to be a non-convex problem. Therefore, we propose two iterative algorithms to obtain the optimum value for the UL sum-rate. The contribution and uniqueness of this paper are as follows.
\begin{itemize}
\item First, the UL weighted sum-rate is maximized for two groups of multi-user MUs based on channel assignment, and the time resource and power resource allocations. Unlike the H-AP architectures presented in \cite{Ju14,Hoon016,Yu19} that considered single antennas for WPT DL and WIT UL, this work considers a H-AP operating in a FD mode with multiple antennas for WPT DL and WIT UL communication between the H-AP and the multi-user MUs. Also, the system models presented in \cite{Ju14,Hoon016,Yu19} operate in TDMA mode. However, the system model studied in this work focuses on a TDD mode of operation. The non-linear EH model is also investigated in this paper\footnote{The EH harvesting models used in the reviewed literature considered linear EH models \cite{Ju14,Hoon016,Yu19,Wei19,Ma19,Li19}. Recently, the non-linear EH model was presented to mimic realistic/real-world EH at an EH node/device, hence, our consideration of the non-linear EH model. \cite{Wei19,Ma19,Li19}.}. Another difference between this investigation and that of the works in \cite{Ju14,Hoon016,Yu19,Shi19,Nguyen19,Solanki19} lies with how the MUs communicate with the H-AP. This paper considers a scenario where some MUs operate in DL and others operate in UL over the same frequency band at the same time instance. However, the works in \cite{Ju14,Hoon016,Yu19} considered the scenario where MUs uniformly perform UL and DL communication together at the same time instance. The distinguishing factor between this work and that presented in \cite{Duy16} lies in the application of WPT (i.e., WPCN) and the consideration of only UL weighted sum-rate maximization in this system model. In contrast, the work in \cite{Duy16} is a non-WPT with UL and DL weighted sum-rate maximization.
\item To add on to perfect CSI scenarios considered in \cite{Ju14,Hoon016,Yu19,Duy16}, the effects of imperfect CSI at the FD H-AP on the UL sum-rate and the MUs' EH is investigated in this paper. This consideration is influenced by real-world scenarios where the H-AP may not be able to perfectly estimate the channel gains of communicating nodes \cite{Yoo06,Xiang12}.
\item From the UL weighted sum-rate problem, the joint optimization problem is found to be non-convex with respect to the time, power, and channel allocation variables. Hence, two algorithms are proposed to find the maximum UL weighted sum-rate. The first algorithm is a fixed time resource scheme in which the time resource is assigned a fixed value and not optimized. With the second proposed scheme, the time resource is optimized in addition to the other variable. Hence, the second scheme is the optimal scheme that achieves the maximum UL weighted sum-rate. 
\item Finally, the FD system model and the proposed resource allocation schemes to the HD-WPCN system model. The HD-WPCN is compared to the FD-WPCN in terms of QoS (i.e., sum-rate), power resource (i.e., the power resources of the AP and MU nodes), and the computational complexity of implementing the proposed algorithm.
\end{itemize}

From the above literature review, it is apparent a lot of research, and innovative devices (e.g., wi-charge technology, Pi charger, energous RF chargers and Warp RF wireless charging systems) for WPT are under study or development. However, most RF WPT research and development projects focus on using the full-band for RF WPT. This implies that, for the development and implementation of both information and power transmit-and-receive devices and systems, signal transmission must use either out-of-band FD or HD method. Therefore, for effective use of the frequency band, in-band FD method must be used. This is the main advantage of this paper concerning both information and power transfer. Finally, the presented system model and proposed schemes in this paper can be used in different scenarios/situations of wireless communication. For example, it can be implemented in a wireless sensor network data collection system. Here, a deployed sensor node could be in one of two modes. The first is the sleep/charging mode in which the sensor node is wirelessly charging its battery and collecting data (e.g., temperature, pressure, humidity, moisture). The other is the wake-up/active mode in which the senor node transmit their data to a receiver. Each mode would occur either during the first or second time slots. With this example, the energy-transmitter and information-receiver could be a gateway within the network serving/connecting two sets of wireless sensor networks communicating with a particular central system.
 
The rest of the paper is organized as follows: Section \ref{secsystem_model} contains the presentation of the system model and the formulation of the sum-rate maximization problem. The proposed optimal scheme is presented in Section \ref{secoptimaldesign}. The special case of a HD system is briefly discussed in Section \ref{secspeccase}. Simulation results based on the proposed optimum sum-rate solution are provided in Section \ref{secresults}. Finally, concluding remarks on this research work are provided in Section \ref{secconclusion}.

\emph{Notations:} $\mathbf{A}$, $\mathbf{a}$, and $a$ denote a matrix, vector, and a scalar variable, respectively. $\mathbf{A} \in \mathbb{C}^{M\times N}$ is a matrix with dimensions $M$ by $N$, and $\mathbf{a} \in \mathbb{C}^{M}$ is a vector with dimensions $M$ by $1$. $\mathbf{I}_{M}$ represents an identity matrix of dimension $M$ by $M$. Also, let $a\sim\mathcal{CN}(0,\sigma^{2})$ denote circularly symmetric complex Gaussian random variable, a, with zero mean and variance of $\sigma^{2}$. Some of the significant abbreviations and notations used in this paper are summarized in Tables \ref{tababbretable} and \ref{tabvariatable}, respectively.

\begin{table}[t!]
\caption{Summary of Abbreviations}
\label{tababbretable}
\begin{tabular}{l||l}
\hline
\bf{Acronym} & \bf{Definition}                                      \\ \hline\hline
AP           & Access Point                                         \\ \hline
AWGN         & Additive White Gaussian Noise                        \\ \hline
BS           & Base Station                                         \\ \hline
CSI          & Channel State Information                            \\ \hline
DL           & Downlink                                             \\ \hline  
EH           & Energy Harvesting                                    \\ \hline
FD           & Full-Duplex                                          \\ \hline
GLSM         & Golden Line Search Method                            \\ \hline 
HD           & Half-Duplex                                          \\ \hline
H-AP         & Hybrid Access Point                                  \\ \hline
IoT          & Internet-of-Things                                   \\ \hline
MU           & Mobile User                                          \\ \hline
MIMO         & Multiple Input Multiple Output                       \\ \hline
PS           & Power Splitting                                      \\ \hline
QoS          & Quality-of-service                                   \\ \hline
RF           & Radio Frequency                                      \\ \hline
RSI          & Residual Self-Interference                           \\ \hline
SI           & Self-Interference                                    \\ \hline
SIC          & Self-Interference Cancellation                       \\ \hline
SWIPT        & Simultaneous Wireless Information and Power Transfer \\ \hline
SNR          & Signal-to-Noise Ratio                                \\ \hline
TS           & Time Switching                                       \\ \hline
TDD          & Time Division Duplex                                 \\ \hline
TDMA         & Time-Division-Multiple-Access                        \\ \hline
UL           & Uplink                                               \\ \hline
WMMSE        & Weighted Minimum Mean Square Error                   \\ \hline
WIT          & Wireless Information Transfer                        \\ \hline
WPT          & Wireless Power Transfer                              \\ \hline
WIT          & Wireless Information and Power Transfer              \\ \hline
WPCN         & Wireless Powered Communication Network               \\ \hline\hline
\end{tabular}
\end{table}

\section{System Models and Problem Formulations}
\label{secsystem_model}
As mentioned in the introduction, this paper focuses on a wireless communication system model where MUs communicate with a H-AP (i.e., the H-AP serves as both the gateway and centralized system connected to the internet for the considered IoT network). The MUs possess single antennas, while the H-AP is equipped with multiple antennas. The H-AP operates in the FD mode, while the MUs operate in the HD mode. In this work, it is assumed that the MUs are sensor nodes, hence operate in HD mode due to their limitations in circuitry and size\cite{Duy16}. This implies that each MU has two orthogonal channels it accesses for their DL and UL communications. Communication between the H-AP and the MUs are in TDD mode. Hence, communication occurs in two phases (i.e., two subsequent time slots). In this work, the assumption that the H-AP knows all the channels used for communication through channel estimation is made\footnote{The CSI for the multi-user D2D sensor network can be acquired during the training phase for channel gain estimation between the MUs and the H-AP (i.e., gateway and central system unit) \cite{Zhao04,Taricco12,Wang18}.}. However, the possibility exists that the channel knowledge is imperfect, leading to the channel estimation error in practical systems. In the FD operation of the H-AP, it beamforms energy signals to the MUs and receives information signals from the MUs at the same time. Due to the FD operation of the H-AP, it experiences SI at its receivers from its transmitters. Hence, the H-AP undergoes SIC, but since the SIC process is not perfect, there will be some residual self-interference (RSI) signal within the H-AP's received signal. 

\begin{figure}[t!]
\centering
\includegraphics[width=3.2in]{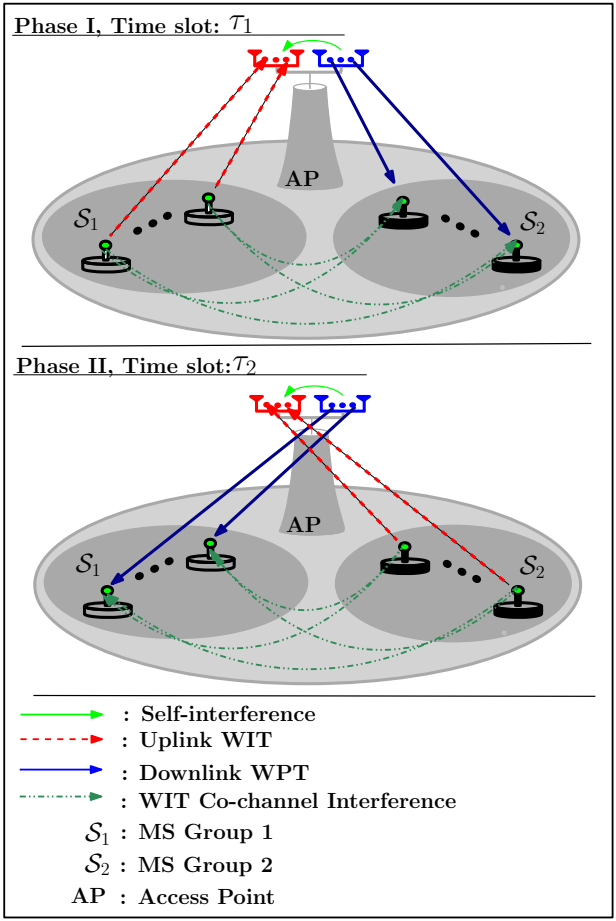}
\caption{System model for the proposed multiuser FD-WPCN.}
\label{figFD-SWIPTModel}
\end{figure}

The MUs harvest energy during their DL communication with the H-AP, and transmit information to the H-AP for their UL communication. In the first phase that lasts $\tau_{1}$ seconds, the MUs are grouped into two based on whether a MU is communicating in WIT UL or EH DL. Please note that there is no specific predetermined UL or DL channel assignment for the MUs. For example, using Fig. \ref{figFD-SWIPTModel}, consider six MUs communicating with the H-AP, where the $3$ MUs in group $\mathcal{S}_1$ receive DL energy signals from the H-AP while the other $3$ MUs in group $\mathcal{S}_2$ transmit UL information signals to the H-AP. Then, in the second phase that lasts $\tau_2$ seconds, the roles of groups $\mathcal{S}_1$ and $\mathcal{S}_2$ are switched. However, the number of MUs in each group may change depending on whether a node switches from EH to WIT. That is, at another time instance, group $\mathcal{S}_1$ may have $2$ MUs communicating in DL while group $\mathcal{S}_2$ has $4$ MUs communicating in UL with the H-AP. The grouping depends on what form (i.e., WET or WIT) of communication an MU is performing in each time instance. This switching between MU groups by MUs is determined by a factor/variable called the channel allocation factor in this work. 
\begin{table}[t!]
\caption{Summary of Variables and Symbols}
\label{tabvariatable}
\begin{tabular}{l||l}
\hline
\bf{Variable}                                      & \bf{Definition}                                                                              \\ \hline\hline
$\tau_{1}$                                         & First time slot                                                                              \\ \hline
$\tau_{2}$                                         & Second time slot                                                                             \\ \hline
$\mathcal{S}_{1}$                                  & Group one MUs operate in DL mode during the $\tau_{1}$                                       \\ \hline
$\mathcal{S}_{2}$                                  & Group two MUs operate in DL mode during the $\tau_{2}$                                       \\ \hline
$\mathcal{C}_{1}$                                  & Channel accessed during the $\tau_{1}$                                                       \\ \hline
$\mathcal{C}_{2}$                                  & Channel accessed during the $\tau_{2}$                                                       \\ \hline
$l$, $\hat{l}$                                     & $\mathcal{S}_{l}$, $\mathcal{C}_{l}$, and $\tau_{l}$ index, where $l=1,2$ and $\hat{l}=3-l$  \\ \hline
MS$_{k}$                                           & The $k$-th MU                                                                                \\ \hline
$\mathbf{h}_{k,l}\in \mathbb{C}^{M\times 1}$       & DL access channel representation for MU$_{k}$ during $\tau_{l}$                              \\ \hline
$\mathbf{g}_{k,l}\in \mathbb{C}^{M\times 1}$       & UL access channel representation for MU$_{k}$ during $\tau_{l}$                              \\ \hline
$\hat{g}_{kj,l}$                                   & MU$_{k}$ and MU$_{j}$ co-channel interference during $\tau_{l}$                              \\ \hline
$\hat{\mathbf{H}}_{l} \in \mathbb{C}^{M\times M}$  & H-AP loop-back channel during $\tau_{l}$                                                      \\ \hline
$M$                                                & A set of antennas at the H-AP used for either UL/DL                                          \\ \hline\hline
\end{tabular}
\end{table}

The time allocation indexes $\tau_1$ and $\tau_2$ should be optimized according to the channel conditions to maximize the entire throughput performance\footnote{A major distinguishing point between this work and the work in \cite{Duy16} is that each MU undergoes EH in DL to facilitate its WIT UL communication. Also, this paper considers only UL sum-rate, unlike the work in \cite{Duy16} that considered both UL and DL sum-rate. In addition, this work covers an imperfect CSI estimation scenario, unlike the conference version of this paper that considered perfect CSI estimation \cite{Asiedu18}.}. The H-AP has $2M$ antennas where the first $M$ antennas are for WIT UL, while the rest are for WPT DL. In this paper, $\mathbf{h}_{k,l}$ and $\mathbf{g}_{k,l}$ are the estimated DL and UL channels of MU$_k$, respectively, where $k=1,2,\ldots,K$ and $l$ represents the phase number, that is, $l=1$ and $2$, indicating phase $1$ and $2$ operations, respectively. The loop-back channel at the H-AP on phase $l$ is denoted as $\hat{\mathbf{H}}_l \in \mathbb{C}^{M\times M}$ having  independent and identically distributed (i.i.d.) Gaussian components with zero mean and variance $\sigma^{2}_{\hat{H}}$. The co-channel interaction between MU$_k$ and MU$_j$ on phase $l$ is denoted by $\hat{g}_{kj,l}$. This work assumes the CSI is estimated with a processing unit at the H-AP. 

The imperfect channel estimation at the H-AP is modeled as
\begin{equation}
\begin{aligned}
&\quad \mathbf{h}_{k,l}=\mathbf{\hat{h}}_{k,l}+\mathbf{\tilde{h}}_{k,l},\text{ }\text{ }\text{ }\text{ }\mathbf{g}_{k,l}=\mathbf{\hat{g}}_{k,l}+\mathbf{\tilde{g}}_{k,l},
\end{aligned}
\end{equation}
where $\mathbf{\hat{h}}_{k,l}$ and $\mathbf{\hat{g}}_{k,l}$ are the estimated channels. $\mathbf{\tilde{h}}_{k,l}$ and $\mathbf{\tilde{g}}_{k,l}$ are the channel estimation errors. Each vector component of the channel estimation error is modeled as a zero mean complex Gaussian random variable with variance $\sigma^{2}_{\mathbf{h}_{k,l}}$ and $\sigma^{2}_{\mathbf{g}_{k,l}}$, respectively. Since, the system model under consideration operates in TDD mode,  channel reciprocity is assumed in this paper. Hence, $\mathbf{g}_{k,l}=\mathbf{h}^{*}_{k,l}$, and further assume that $\sigma^{2}_{\mathbf{h}_{k,l}}=\sigma^{2}_{\mathbf{g}_{k,l}}$. Also, the estimation error variance is fixed as $\sigma^{2}_{\mathbf{E}}$. $\sigma^{2}_{\mathbf{E}}$ can be determined through various methods depending on the channel dynamics and channel estimation schemes \cite{Yoo06,Xiang12}. Channel $\mathbf{\hat{h}}_{k,l}=[\hat{h}_{1,k,l},\hat{h}_{2,k,l},\ldots,\hat{h}_{M,k,l}]$ has a dimension of $M\times 1$, where each component is modeled as $\hat{h}_{i,k,l}=C_{0}d^{-\varepsilon_{h}/2}\zeta_{i}$, where $i=1,2,\ldots,M$, and $\mathbf{\hat{g}}_{k,l}=\mathbf{\hat{h}}^{*}_{k,l}$. $\varepsilon_{h}=3$ is the pathloss exponent, $C_{0}$ is the attenuation co-efficient at reference distance of $1$m and $\zeta_{i}\sim\mathcal{CN}(0,1-\sigma^{2}_{\mathbf{E}})$. Also, the erroneous channel component is modeled as $\tilde{h}_{i,k,l}=C_{0}d^{-\varepsilon_{h}/2}\tilde{\zeta}_{i}$ with $\tilde{\zeta}_{i}\sim\mathcal{CN}(0,\sigma^{2}_{\mathbf{E}})$. However, for the work presented in this paper, it is assumed that $\mathbf{\hat{h}}_{k,l}$ and $\sigma^{2}_{\mathbf{E}}$ are already known\footnote{The main goal of this paper is to study the effect imperfect CSI has on the system performance, and not the estimation error calculation. Hence, channel estimation is not covered.}. 

During DL communication, the received energy signal from the H-AP at MU$_k$ over phase $l$ is given as
\begin{equation}
\label{equmyequ1WPCN}
\begin{aligned}
y^{DL}_{k,l}=\underbrace{\mathbf{h}^{H}_{k,l}\sum_{j\in \mathcal{S}_l}\mathbf{w}_{j,l}s^{DL}_{j}}_{\text{Energy signal}}+\underbrace{\sum_{j\in \mathcal{S}_{\hat{l}}}\hat{g}_{kj,l}\sqrt{P^{UL}_{j,l}}s^{UL}_{j}}_{\text{UL co-channel signal}}+z^{DL}_{k,l},
\end{aligned}
\end{equation}
where $s^{DL}_{k}$ is the energy signal transmitted from the H-AP and $\mathbf{w}_{k,l} \in \mathbb{C}^{K}$ denotes DL beamformer on phase $l$. $P^{UL}_{j,l}$ is the transmit power of MU$_j$ and $z^{DL}_{k,l}$ is the additive white Gaussian noise (AWGN) at MU$_k$ with variance $\sigma^{2}_{k,l}$. Here, the assumption that the UL co-channel signal and noise EH components in (\ref{equmyequ1WPCN}) are relatively negligible compared to the actual transmitted energy signal is made. Thus, from (\ref{equmyequ1WPCN}) and the EH model presented in \cite{Shi19,Nguyen19,Solanki19}, the harvested energy at MU$_{k}$ is given as\footnote{Please note that the MUs harvest energy using the actual channel, $h_{k,l}$, and not the estimated channel, $\hat{h}_{k,l}$. This is because the EH harvesting phase at the MUs do not need CSI in their process. Each MU just harvests energy from the signal it receives. However, during signal transmission from the H-AP and MUs, the CSI is needed. Hence, $\hat{h}_{k,l}$ is used by both the H-AP and MUs.} 
\begin{equation}
Q_{k,l}=\tau_{l}\beta_{k}\text{min}\Big(\sum_{j\in \mathcal{S}_l}\vert\mathbf{h}^{H}_{k,l}\mathbf{w}_{j,l}\vert^2,P_{TH}\Big),
\end{equation} 
where $\beta_{k}$ and $P_{TH}$ are the EH efficiency and the saturation threshold of EH receiver at MU$_k$, respectively. Now, the power facilitating MU$_k$'s WIT UL communication in the next phase is represented as 
\begin{equation}
\begin{aligned}
&\!\!\!\!\hat{Q}_{k,\hat{l}}=\frac{Q_{k,l}}{\tau_{\hat{l}}}=\frac{\tau_{l}}{\tau_{\hat{l}}}\beta_{k}\text{min}\Big(\sum_{j\in \mathcal{S}_l}\vert\mathbf{h}^{H}_{k,l}\mathbf{w}_{j,l}\vert^2,P_{TH}\Big),
\end{aligned}
\end{equation}
where $\hat{l}=3-l$ that is complement to $l$. 

For UL communication on phase $\hat{l}$, the received signal at H-AP for the UL MUs is written as
\begin{equation}
y^{UL}_{\hat{l}}=\underbrace{\sum_{k\in \mathcal{S}_l}\mathbf{\hat{g}}_{k,\hat{l}}\sqrt{P^{UL}_{k,\hat{l}}}s^{UL}_{k}}_{\text{signal from }\mathcal{S}_{l}\text{ MUs}}+\underbrace{\hat{\mathbf{H}}_{\hat{l}}\mathbf{x}^{DL}_{\hat{l}}}_{\text{residual SI on } \mathcal{C}_{\hat{l}}}+z^{UL}_{\hat{l}},
\end{equation} 
where $\hat{\mathbf{H}}_{\hat{l}}\mathbf{x}^{DL}_{\hat{l}}$ is the RSI and $\mathbf{x}^{DL}$ is the energy signal being transmitted to the EH MUs in this time slot. The H-AP uses a receive filter $\mathbf{v}_{k,\hat{l}}$ to detect the received signal from MU$_k$, which is expressed as
\begin{equation}
\begin{aligned}
&\hat{s}^{UL}_{k}=\mathbf{v}^{H}_{k,\hat{l}}\mathbf{\hat{g}}_{k,\hat{l}}\sqrt{P^{UL}_{k,\hat{l}}}s^{UL}_{k}+\mathbf{v}^{H}_{k,\hat{l}}\sum_{j\in \mathcal{S}_l,j\neq k}\mathbf{\hat{g}}_{j,\hat{l}}\sqrt{P^{UL}_{j,\hat{l}}}s^{UL}_{j}\\& \text{ }
\begin{aligned}
& && & && & &+\mathbf{v}^{H}_{k,\hat{l}}\hat{\mathbf{H}}_{\hat{l}}\sum_{j\in \mathcal{S}_{\hat{l}}}\mathbf{w}_{j,\hat{c}}s^{DL}_{j}+\mathbf{v}^{H}_{k,\hat{l}}z^{UL}_{\hat{l}}.
\end{aligned}
\end{aligned}
\end{equation}
 The UL signal-to-noise ratio (SNR) for MU$_k$ in phase $\hat{l}$ is expressed as
\begin{equation}
\label{equSNRWPCN}
\gamma^{UL}_{k,\hat{l}}=\frac{P^{UL}_{k,\hat{l}}\vert\mathbf{v}^{H}_{k,\hat{l}}\mathbf{\hat{g}}_{k,\hat{l}}\vert^2}{\sum_{j\in \mathcal{S}_l,j\neq k}P^{UL}_{j,\hat{l}}\vert\mathbf{v}^{H}_{k,\hat{l}}\mathbf{\hat{g}}_{j,\hat{l}}\vert^2+RSI_{k,\hat{l}}+\sigma^{2}_{\hat{l}}\Vert\mathbf{v}^{H}_{k,\hat{l}}\Vert^{2}}.
\end{equation} 
where $RSI_{k,\hat{l}}=\hat{\sigma}^{2}_{\hat{H}}\Vert\mathbf{v}_{k,\hat{l}}\Vert^2\sum_{j\in \mathcal{S}_{\hat{l}}}\Vert\mathbf{w}_{j,\hat{l}}\Vert^2,$ represents the average power of RSI.
 
The rate for the UL communication of MU$_k$ is given as
\begin{equation}
\label{equratefind}
R^{U}_{k}=\sum^{2}_{l=1}a_{k,\hat{l}}\tau_{l}\log_2(1+\gamma^{UL}_{k,l}), 
\end{equation} 
where $a_{k,l}$ is a binary variable indicating the association of MU$_k$ to a particular MU group, $\mathcal{S}_l$ \cite{Duy16}. For example, if $a_{k,l}=1$, then MU$_k$ undergoes EH in $\tau_{1}$ and belongs to group $\mathcal{S}_1$. To jointly optimize sum-rate with respect to  $\mathbf{w}_{k,l}$, $\mathbf{v}_{k,l}$, $\tau_{l}$, $a_{k,l}$ and $P^{UL}_{k,\hat{l}}$, the optimization problem  is formulated for the FD-WPCN system as 
\begin{equation}
\label{equoptmainWPCN}
\begin{aligned}
& \underset{\tau_{l},a_{k,l},\mathbf{w}_{k,l},\mathbf{v}_{k,l},P^{UL}_{k,\hat{l}}}{\text{maximize}}
\text{ }\sum^{K}_{k=1}R^{U}_{k} \\
& \text{subject to} 
\begin{aligned}  
& &  \sum^{K}_{k=1}\sum^{2}_{l=1}\Vert\mathbf{w}_{k,l}\Vert^2\leq P^{DL}_{0,max},
\end{aligned}\\
& \text{}   
\begin{aligned}  
& & & & & & & & & & & P^{UL}_{k,\hat{l}} \leq \hat{Q}_{k,\hat{l}};\text{ }\forall k,
\end{aligned}\\
& \text{}   
\begin{aligned}  
& & & & & & & & & & & a_{k,l} \in \left\lbrace 0,1\right\rbrace ,\text{ }a_{k,1}+ a_{k,2} =1;\text{ }\forall k,
\end{aligned}\\
& \text{}   
\begin{aligned}  
& & & & & & & & & & & 0 < \tau_{l} < 1,\text{ }\tau_{1}+ \tau_{2} =1,
\end{aligned}
\end{aligned}
\end{equation}
where $P^{DL}_{0,max}$ is the transmit power at the H-AP. From problem (\ref{equoptmainWPCN}), the first constraint is the DL transmit power budget. The DL power budget means the total power allocation for the energy signal beamforming cannot exceed the maximum available power at the H-AP. Each MU's UL power budget is also defined as the second constraint in the problem (\ref{equoptmainWPCN}). The power used for information signal transmission during the UL communication should also not exceed the amount of harvested energy at a MU. Each MU can be assigned to one group. The channel assignment binary variable determines which group a MU belongs to at a given time. For example, if MU$_{k}$ is assigned to $\mathbf{h}_{k,1}$, then the MU$_{k}$ undergoes DL transmission in channel $\mathbf{h}_{k,1}$ (i.e., $a_{k,1}=1$ and $a_{k,2}=0$). Whereas, MU$_{k}$ assigned to $\mathbf{g}_{k,1}$ performs UL in that channel (i.e., $a_{k,1}=0$ and $a_{k,2}=1$). The sum of the channel assignment for a MU should be one, as seen in the third constraint. Finally, the time allocation constraint for communication over the two channels is defined in the final constraint. The total time used to access both channels by the MUs should be equal to the total time for communication. The total time for communication is assumed to be unity. In the next section, the solution to the sum-rate maximization problem in (\ref{equoptmainWPCN}) is presented. Note that problem (\ref{equoptmainWPCN}) is a non-convex mixed-integer program due to variables $a_{k,l}$ and $a_{k,2}$.
\section{Problem Optimization}
\label{secoptimaldesign}
Problem (\ref{equoptmainWPCN}) is non-convex for the joint optimization of all variables \cite{Duy16}. To acquire the optimal solution,  (\ref{equoptmainWPCN}) is solved via the alternative optimization process. Individually, one variable is evaluated at a time while keeping the other variables constant in each iterative step. The solution to the DL energy beamformer is obtained first by using the Lagrange duality method approach in subsection \ref{secBSbeamtransWPCN} as sub-problem $1$. Next, the optimal receiver filter and the UL MS transmit power are computed by converting the sum-rate maximization problem to a weighted minimum mean square error (WMMSE) minimization problem as sub-problem $2$ in subsection \ref{subsecWMMSEminWPCN}. The channel allocation is then solved using a mixed-integer linear programming approach as sub-problem $3$ in subsection \ref{secchanallWPCN}. Finally, optimal time allocation is obtained using sub-problem $4$, which is a sum-rate maximization problem for only the time resource in subsection \ref{secTimeslotWPCN}.
\subsection{BS transmit beamforming optimization}
\label{secBSbeamtransWPCN}
To solve for the energy beamformer, the Lagrangian of problem (\ref{equoptmainWPCN}) with respect to all variables and constraints is given as 
\begin{equation}
\label{equlagranit}
\begin{aligned}
&\mathcal{L}(\tau_{l},a_{k,l},\mathbf{v}_{k,l},\mathbf{w}_{k,l},P^{UL}_{k,l},\lambda^{DL},\lambda^{UL}_{k})\\&\quad=\sum^{K}_{k=1}\sum^{2}_{l=1}a_{k,\hat{l}}\tau_{l}\log_2(1+\gamma^{UL}_{k,l})+\lambda^{DL} P^{DL}_{0,max}\\ & \text{ }
\begin{aligned}
&&-\sum^{K}_{k=1}\sum^{2}_{l=1}\lambda^{UL}_{k}P^{UL}_{k,l}- \lambda^{DL}\sum^{K}_{k=1}\sum^{2}_{l=1}\Vert\mathbf{w}_{k,l}\Vert^{2}  
\end{aligned}
 \\ & \text{ }
\begin{aligned}
&&+\sum^{K}_{k=1}\sum^{2}_{l=1}\lambda^{UL}_{k}\beta_{k}\frac{\tau_{l}}{\tau_{\hat{l}}}\sum_{j\in \mathcal{S}_l}\vert\mathbf{h}_{k,l}^{H} \mathbf{w}_{j,l}\vert^{2}.
\end{aligned}
\end{aligned}
\end{equation}
From the above Lagrangian, a theorem for finding the optimal H-AP energy beamforming vector is proposed. Theorem \ref{theoremtheorem1} presented below is used to find the optimal $\mathbf{w}_{k,l}$. This theorem is based on the approach of solving for only $\mathbf{w}_{k,l}$ from the above Lagrangian while treating the other dependent variables of problem (\ref{equoptmainWPCN}) as constants. 
\begin{theorem} \label{theoremtheorem1}
The optimal energy beamformer, $\mathbf{w}_{k,l}$, for the optimization problem (\ref{equoptmainWPCN}) is expressed as
\begin{equation}
\label{equmyfindwWPCN}
\mathbf{w}_{k,l}^{\star}=\sqrt{P^{DL}_{0,max}}\mathbf{u}_{B,1},
\end{equation}
where $\mathbf{u}_{B,1}$ in equation (\ref{equmyfindwWPCN}) is the unit-norm eigenvector corresponding to the maximum eigenvalue of matrix $B\triangleq\frac{\tau_{l}}{\tau_{\hat{l}}} \sum_{j\in \mathcal{S}_l}\lambda^{UL}_{j}\beta_{j}\Big[\mathbf{\hat{h}}_{j,l}\mathbf{\hat{h}}_{j,l}^{H}\Big]$\footnote{In this work, the H-AP runs the proposed algorithm. Hence, $\mathbf{u}_{B,1}$ is calculated using the estimated channel, $\hat{h}_{k,l}$, and not the actual channel. Therefore, $\tilde{h}_{j,l}$ is not used to determine $\mathbf{u}_{B,1}$.}.
\end{theorem}

The optimal $\mathbf{w}_{k,l}$ solution presented in Theorem \ref{theoremtheorem1}, is obtained by using the differential of (\ref{equlagranit}) and the Karush–Kuhn–Tucker (KKT) conditions with respect to $\mathbf{w}_{k,l}$. From the differential equation, it is established that the optimal $\mathbf{w}_{k,l}$ can be found from eignvalue decomposition. 

\textit{Proof:} Please, see Appendix \ref{AppApp1} for the detailed derivation of the optimal $\mathbf{w}_{k,l}$. 

Note that the solution acquired for $\mathbf{w}_{k,l}$ is dependent on finding the optimal $\lambda^{UL}_{j}$. Next, closed-form solutions for the receiver filter and the MS UL transmit power are presented. From problem (\ref{equoptmainWPCN}), it is very difficult to solve for the two variables. Hence, the sum-rate optimization problem is converted to a WMMSE minimization problem. Using the WMMSE minimization problem, solutions for the receiver filter and the MU UL transmission power are deduced in the next subsection.
\subsection{Equivalent WMMSE minimization}
\label{subsecWMMSEminWPCN}
By leveraging the equivalence relationship between sum-rate maximization and WMMSE minimization as shown in \cite{Christensen08}, the sum-rate problem in (\ref{equoptmainWPCN}) can be reformulated to a WMMSE minimization problem by considering the variables $P^{UL}_{k,l}$ and $\mathbf{v}_{k,l}$, and the $P^{UL}_{k,l}$ constraint as 
\begin{equation}
\label{equWMMSEWPCN}
\begin{aligned}
& \underset{\mathbf{v}_{k,l},\vartheta^{UL}_{k,l},P^{UL}_{k,l}}{\text{minimize}}\text{ }\Gamma^{U}\\
& \text{subject to}   
\begin{aligned}  
& & P^{UL}_{k,l} \leq \hat{Q}_{k,l};\text{ }\forall k,
\end{aligned}
\end{aligned}
\end{equation}
where $\vartheta^{UL}_{k,l}$ is the UL MMSE weight for MU$_k$ on channel $\mathcal{C}_{l}$,
\begin{equation}
\Gamma^{U}=\sum^{K}_{k=1}\sum^{2}_{l=1}a_{k,\hat{l}}\left(\vartheta^{UL}_{k,l}e^{UL}_{k,l}-\log \vartheta^{UL}_{k,l}\right),
\end{equation}
and
\begin{equation}
\label{equedlWPCN}
\begin{aligned}
& e^{UL}_{k,l}=\vert 1-\sqrt{P^{UL}_{k,l}}\mathbf{v}^{H}_{k,l}\mathbf{\hat{g}}_{k,l}\vert^{2}+\sum_{j\in \mathcal{S}_{\hat{l}},j\neq k}P^{UL}_{j,l}\vert\mathbf{v}^{H}_{k,l}\mathbf{\hat{g}}_{j,l}\vert^{2}\\ & \text{ }
\begin{aligned}
&&&&&&&&&+\Big( \sum_{j\in \mathcal{S}_{l}}\Vert\mathbf{w}_{j,l}\Vert^{2}\hat{\sigma}^{2}_{\hat{H}}+\sigma^{2}_{l}\Big)\Vert\mathbf{v}_{k,l}\Vert^{2}.  
\end{aligned}
\end{aligned}
\end{equation}
Problem (\ref{equWMMSEWPCN}) is a convex joint optimization problem with respect to all the variables \cite{Shi11,Pan15,Tian20}. The proof of convexity can be found in Appendix \ref{AppApp2a}.
The optimal receiver filter, $\mathbf{v}_{k,l}$ is deduced from
\begin{equation}
\begin{aligned}
\mathbf{v}^{\star}_{k,l}=\arg \underset{\mathbf{v}_{k,l}}{\text{min}} &  &  e^{UL}_{k,l}
\end{aligned}
\end{equation} 
as
\begin{equation}
\label{equTbeamWPCN}
\mathbf{v}^{\star}_{k,l}=\Big(\sum_{j\in \mathcal{S}_{\hat{l}}}P^{UL}_{j,l}\mathbf{\hat{g}}_{j,l}\mathbf{\hat{g}}^{H}_{j,l}+ c\mathbf{I}_{M} \Big)^{-1}\sqrt{P^{UL}_{k,l}}\mathbf{\hat{g}}_{k,l}, 
\end{equation} 
where
\begin{equation}
c=\Big( \sum_{j\in \mathcal{S}_{l}}\Vert\mathbf{w}_{j,l}\Vert^{2}\hat{\sigma}^{2}_{\hat{H}}+\sigma^{2}_{l}\Big).
\end{equation} 
The above solution is acquired from the differentiation of $e^{UL}_{k,l}$ with respect to $\mathbf{v}_{k,l}$, then equated to zero, and finally solved to find $\mathbf{v}_{k,l}$. 
The optimal UL weight, $\vartheta^{UL\star}_{k,l}$, is estimated as 
\begin{equation}
\label{equULWWPCN}
\vartheta^{UL\star}_{k,l}=\frac{1}{1-\sqrt{P^{UL}_{k,l}}\mathbf{v}^{H}_{k,l}\mathbf{\hat{g}}_{k,l}}
\end{equation}
from the differential of 
\begin{equation}
\begin{aligned}
\vartheta^{UL\star}_{k,l}=\arg \underset{\vartheta^{UL}_{k,l}}{\text{min}} &  &  \left[ \vartheta^{UL}_{k,l}e^{UL}_{k,l}-\log \vartheta^{UL}_{k,l}\right] 
\end{aligned}
\end{equation}
being equated to zero, and solving for $\vartheta^{UL\star}_{k,l}$.
Now, simplifying and grouping like terms in $\Gamma^{U}$ of problem (\ref{equWMMSEWPCN}) produces 
\begin{equation}
\begin{aligned}
&\sum^{K}_{k=1}\sum^{2}_{l=1}\bigg[a_{k,\hat{l}}\vartheta^{UL}_{k,l}\Vert\mathbf{v}_{j,l}\Vert^{2}\Big(\sum_{j\in \mathcal{S}_{l}}\Vert\mathbf{w}_{j,l}\Vert^{2}\hat{\sigma}^{2}_{\hat{H}}+\sigma^{2}_{l}\Big)\\ & \text{ }
\begin{aligned}
&&&&&&+a_{k,\hat{l}}(\vartheta^{UL}_{k,l}-\log \vartheta^{UL}_{k,l})\bigg]
\end{aligned}\\ & \text{ }
\begin{aligned}
&+\sum^{K}_{k=1}\sum^{2}_{l=1}\bigg[a_{k,\hat{l}}\vartheta^{UL}_{k,l}\sum_{j\in \mathcal{S}_{\hat{l}}}P^{UL}_{j,l}\vert\mathbf{v}^{H}_{k,l}\mathbf{\hat{g}}_{j,l}\vert^{2}
\end{aligned}\\ & \text{ }
\begin{aligned}
&&&&&&&&-\sqrt{P^{UL}_{k,l}}a_{k,\hat{l}}\vartheta^{UL}_{k,l}(\mathbf{v}^{H}_{k,l}\mathbf{\hat{g}}_{k,l}+\mathbf{\hat{g}}^{H}_{k,l}\mathbf{v}_{j,l})\bigg].
\end{aligned}
\end{aligned}
\end{equation}
This is done to identify the parts of the problem (\ref{equWMMSEWPCN}) influenced by $P^{UL}_{k,\hat{l}}$ in order to aid with the optimization. To obtain the optimal MU$_{k}$ transmit power, let us denote $\Phi$ as
\begin{equation}
\begin{aligned}
&\Phi=\sum^{K}_{k=1}\sum^{2}_{l=1}\bigg[ a_{k,\hat{l}}\vartheta^{UL}_{k,l}\sum_{j\in \mathcal{S}_{\hat{l}}}P^{UL}_{j,l}\vert\mathbf{v}^{H}_{k,l}\mathbf{\hat{g}}_{j,l}\vert^{2}\\ & \text{ }
\begin{aligned}
&&&&&&&&&-\sqrt{P^{UL}_{k,l}}a_{k,\hat{l}}\vartheta^{UL}_{k,l}(\mathbf{v}^{H}_{k,l}\mathbf{\hat{g}}_{k,l}+\mathbf{\hat{g}}^{H}_{k,l}\mathbf{v}_{j,l})\bigg]. 
\end{aligned}
\end{aligned}
\end{equation}
The optimization problem for finding the UL transmit power allocation can be expressed as
\begin{equation}
\label{equnewprobWPCN}
\begin{aligned}
& \underset{P^{UL}_{k,\hat{l}}}{\text{minimize}}\text{ }\Phi \\
& \text{subject to} 
\begin{aligned}  
& \text{ }P^{UL}_{k,l}\leq\hat{Q}_{k,l};\text{ }\forall k.
\end{aligned}
\end{aligned}
\end{equation}
The UL transmit power budget is maintained as a constraint for problem (\ref{equnewprobWPCN}), since it considers the $P^{UL}_{k,l}$. By applying the Lagrangian approach to equation (\ref{equnewprobWPCN}), the optimal UL transmit power of MU$_{k}$ is written as 
\begin{equation}
\label{equpowerWPCN}
P^{UL\star}_{k,l}=\Bigg( \frac{a_{k,\hat{l}}\vartheta^{UL}_{k,l}\mathbf{v}^{H}_{k,l}\mathbf{\hat{g}}_{k,l}}{\sum^{\mathcal{S}_{\hat{l}}}_{j=1}a_{j,\hat{l}}\vartheta^{UL}_{j,l}\vert\mathbf{v}^{H}_{j,l}\mathbf{\hat{g}}_{k,l}\vert^{2}-\lambda^{UL}_{k}}\Bigg)^{2}. 
\end{equation}
The solution in equation (\ref{equpowerWPCN}) is acquired by differentiating the Lagrangian of (\ref{equnewprobWPCN}) with respect to $P^{UL}_{k,l}$, and equating the differential to zero. Then, $P^{UL\star}_{k,l}$ is solved by making it the subject. The $\lambda^{UL}_{k}$ can be determined by substituting equation (\ref{equpowerWPCN}) into the MU$_k$ transmit power budget constraint in (\ref{equoptmainWPCN}) and solving for $\lambda^{UL}_{k}$ as 
\begin{equation}
\label{equpowerWPCNlambda}
\lambda^{UL}_{k}=\sum^{\mathcal{S}_{\hat{l}}}_{j=1}a_{j,\hat{l}}\vartheta^{UL}_{j,l}\vert\mathbf{v}^{H}_{j,l}\mathbf{\hat{g}}_{k,l}\vert^{2}-\frac{a_{k,\hat{l}}\vartheta^{UL}_{k,l}\mathbf{v}^{H}_{k,l}\mathbf{\hat{g}}_{k,l}}{\sum^{2}_{l=1}\sqrt{\hat{Q}_{k,l}}}.
\end{equation} 
Next, the solution for the optimal channel assignment together with an iterative algorithm for finding the suboptimal sum-rate is presented  in the next subsection.
\subsection{Channel allocation optimization}
\label{secchanallWPCN}
To obtain the optimal channel assignment, the optimization problem to consider is stated as
\begin{equation}
\label{equoptmain1WPCN}
\begin{aligned}
& \underset{a_{k,l}}{\text{maximize}}
\text{ }\sum^{K}_{k=1}\sum^{2}_{l=1}a_{k,l}\tau_{\hat{l}}\log_{2}(1+\gamma^{UL}_{k,\hat{l}}) \\
& \text{subject to} 
\begin{aligned}  
& &\sum^{2}_{l=1}a_{k,l}=1,\text{ } a_{k,l} \in \{0,1\}.
\end{aligned} 
\end{aligned}
\end{equation}
The above problem is a mixed-integer linear program problem. Hence, it is non-convex but can still be solved using a generic mixed-integer linear program solver. In this work, the branch-and-bound procedure is employed to solve the problem (\ref{equoptmain1WPCN}). This procedure is chosen because the solution sets are binary. In this procedure, the set of all binary combinations of the variable are listed, and the best feasible point is selected. With problem (\ref{equoptmain1WPCN}), the possible feasible sets are $[a_{k,l},a_{k,\hat{l}}]=([0,1],[1,0])$. Both possible sets are used to solve the objective function, and the set with the highest object function value is chosen. Please note that this procedure is repeated for all MUs within the system. Problem (\ref{equoptmain1WPCN}) is used to assign a MU to a particular group and determine the form of interaction the MU will have with the H-AP in each phase (i.e., time slot). 

With fixed time slots, the sum-rate can be found by iteratively solving for $\mathbf{w}_{k,l}$, $\mathbf{v}_{k,\hat{l}}$, $\vartheta^{UL}_{k,\hat{l}}$, $\lambda^{UL}_{k}$, $P^{UL}_{k,\hat{l}}$ and $a_{k,l}$ using equations (\ref{equmyfindwWPCN}), (\ref{equTbeamWPCN}), (\ref{equULWWPCN}), (\ref{equpowerWPCNlambda}), (\ref{equpowerWPCN}), and by solving problem (\ref{equoptmain1WPCN}), respectively. The iterative processes is repeated until the sum-rate converges. To perform the iterative process, $\lambda^{UL}_{k}$, $P^{UL}_{k,\hat{l}}$ and fixed $\tau_l$ must be initialized. During each variable update in an iteration, the current variable is found using the preceding determined variables updates. For example, if $P^{UL}_{k,\hat{l}}$ is currently being updated, then the previously found values, that is, $\mathbf{w}_{k,l}$, $\mathbf{v}_{k,\hat{l}}$, $\vartheta^{UL}_{k,\hat{l}}$, $\lambda^{UL}_{k}$, and $a_{k,l}$ are used to update $P^{UL}_{k,\hat{l}}$. After updating all variables, the current sum-rate, $\sum^{K}_{k=1}R^{U}_{k}(i)$, is calculated and compared to the previously calculated sum-rate, $\sum^{K}_{k=1}R^{U}_{k}(i-1)$, that is, $\vert\sum^{K}_{k=1}R^{U}_{k}(i)-\sum^{K}_{k=1}R^{U}_{k}(i-1)\vert\leq \epsilon$, where $\epsilon$ is a set tolerance value. If the $\vert\sum^{K}_{k=1}R^{U}_{k}(i)-\sum^{K}_{k=1}R^{U}_{k}(i-1)\vert\leq \epsilon$ condition is met, then the optimal sum-rate is achieved. The sum-rate maximization algorithm with a fixed time slot logical framework is outlined in Algorithm \ref{algMGSAlgorithmWPCN}. It should be noted that the time allocation has not been solved yet. The presented algorithm is a suboptimal solution because the time resource is not optimized. The solution to the optimal time allocation problem is presented in the next subsection. In addition to the optimal time allocation solution, the algorithm for finding the optimum sum-rate in the proposed system model is also introduced.
\begin{algorithm}[t!]
\centering
\caption{Suboptimal FD/HD WPCN: Sum-rate optimization scheme with fixed $\tau_l$}
\label{algMGSAlgorithmWPCN}
\begin{algorithmic} 
\STATE Initialize $\lambda^{UL}_{k}$, $P^{UL}_{k,\hat{l}}$ and fix $\tau_l$
\REPEAT 
\STATE Calculate each $\mathbf{w}_{k,l}$ from (\ref{equmyfindwWPCN})
\STATE Update each $\mathbf{v}_{k,\hat{l}}$ with (\ref{equTbeamWPCN})
\STATE Update each $\vartheta^{UL}_{k,\hat{l}}$ with (\ref{equULWWPCN})
\STATE Update each $\lambda^{UL}_{k}$ with (\ref{equpowerWPCNlambda})
\STATE Update each $P^{UL}_{k,\hat{l}}$ with (\ref{equpowerWPCN})
\STATE Update each $a_{k,l}$ and $a_{k,\hat{l}}$ by solving problem (\ref{equoptmain1WPCN})
\UNTIL{$\sum^{K}_{k=1}R^{U}_{k}$ convergence}
\end{algorithmic}
\end{algorithm} 

\subsection{Time allocation optimization}
\label{secTimeslotWPCN}
The optimization problem for the time allocation is expressed as
\begin{equation}
\label{equoptmain2WPCN}
\begin{aligned}
& \underset{\tau_{l}}{\text{maximize}}
\text{ }\sum^{K}_{k=1}\sum^{2}_{l=1}a_{k,\hat{l}}\tau_{l}\log_{2}(1+\gamma^{UL}_{k,l}) \\
& \text{subject to}
\begin{aligned}  
& & P^{UL}_{k,l} \leq \hat{Q}_{k,l};\text{ }\forall k,\text{ }\sum^{2}_{l=1}\tau_{l}=1,\text{ } 0 < \tau_{l} < 1.
\end{aligned}
\end{aligned}
\end{equation}
Considering the Lagrangian presented in equation (\ref{equlagranit}),  the following lemma defines the optimal values of $\tau_l$ and $\tau_{\hat{l}}$.
\begin{Lemma} \label{TheoremTheorem2}
The optimization problem (\ref{equoptmain2WPCN}) is a convex problem. Thus, a simple line search method can be employed to obtain the optimal $\tau^{\star}_{l}$.
\end{Lemma}

\textit{Proof:} For proof of convexity, refer to Appendix \ref{AppApp2}.

In this paper, the golden line search method (GLSM) is employed to find the optimal $\tau^{\star}_{l}$, but other line search methods can be adopted. Note that this lemma assumes that all other dependent variables are locally optimal, that is, the other variables are predetermined from Algorithm \ref{algMGSAlgorithmWPCN}. $\tau^{\star}_{l}$ is updated using the GLSM, which is an iterative line search method. Hence, with each search step, the sum-rate value used in the GLSM is found using a current $\tau^{\star}_{l}$, and the other resource values (i.e., $\mathbf{w}_{k,l}$, $\mathbf{v}_{k,\hat{l}}$, $\vartheta^{UL}_{k,\hat{l}}$, $\lambda^{UL}_{k}$, $P^{UL}_{k,\hat{l}}$ and $a_{k,l}$) found using Algorithm \ref{algMGSAlgorithmWPCN}. The search processing is repeated until the global optimum value of the sum-rate is found. The algorithm for finding the global optimal sum-rate with $\tau^{\star}_l$ is given in Algorithm \ref{algMGSAlgorithmWPCN1}. 

In summary, the sum-rate problem is non-convex for the joint optimization of all variables (i.e., $\tau_l$, $\mathbf{w}_{k,l}$, $\mathbf{v}_{k,\hat{l}}$, $\vartheta^{UL}_{k,\hat{l}}$, $\lambda^{UL}_{k}$, $P^{UL}_{k,\hat{l}}$ and $a_{k,l}$). The sum-rate curve will consist of several local minimums and a global maximum when plotted\footnote{Please note that the sum-rate is a function of five variable. That is, it has 6-dimensions, hence, it would be hard to plot.}. However, converting to the WMMSE problem makes the problem convex with respect to all variables except $a_{k,l}$, $\mathbf{w}_{k,l}$ and $\tau_l$. Since $a_{k,l}$ is a linear programming problem with solutions consisting of either $0$ or $1$, it does not influence the convexity of the WMMSE problem but just gives information on the MU grouping, therefore, its a passive variable. The optimal value of $\mathbf{w}_{k,l}$ depends on the source power and the unit-norm eigenvector of the matrix $\mathbf{B}(\lambda^{UL}_{k},\mathbf{h}_{k,l})$. This implies that the H-AP transmits the same energy symbol to all users, which is the optimal policy \cite{Hoon16,Hoon15}. Since $\mathbf{w}_{k,l}$ is optimal, and $a_{k,l}$ is binary and passive, using both variables in the WMMSE problem and Algorithm \ref{algMGSAlgorithmWPCN} in an iterative manner leads to a local optimal solution of the sum-rate for fixed $\tau_l$. From Lemma \ref{TheoremTheorem2}, the sum-rate is a convex problem with respect to $\tau_l$. Hence, repeating Algorithm \ref{algMGSAlgorithmWPCN} to find the optimal variables and feeding those values to Algorithm \ref{algMGSAlgorithmWPCN1} leads to a higher sum-rate value compared to the sum-rate acquired with Algorithm \ref{algMGSAlgorithmWPCN}. Thus, the sum-rate finally achieved with Algorithm \ref{algMGSAlgorithmWPCN1} after the iterations terminate at the global optimum. This implies that after the final iteration, the acquired optimal variables globally maximize the sum-rate achieved by the system.
\begin{algorithm}[t!]
\centering
\caption{Optimal FD/HD WPCN: Optimal $\tau_l$ line search algorithm}
\label{algMGSAlgorithmWPCN1}
\begin{algorithmic} 
\STATE Initialize $\tau_l$
\REPEAT 
\STATE Calculate sum-rate from Algorithm \ref{algMGSAlgorithmWPCN} with $\tau_l$
\STATE Update $\tau_l$ with chosen line search method updater
\UNTIL{convergence}
\end{algorithmic}
\end{algorithm}

\section{Half-Duplex Mode of Operation}
\label{secspeccase}
In this section, the above FD system model and algorithms are extended to the case where the H-AP operates in HD mode. In the simulation results and discussion section, the performance of the FD-WPCN mode is compared with that of HD-WPCN mode. For a fair comparison, the FD-WPCN algorithm is modified to suit the HD-WPCN system model \cite{Hoon016,Duy16,Hoon15}. For the HD-WPCN mode of operation, the H-AP operates in HD mode as it communications with the $K$ MUs. This implies that all the $K$ MUs are communicating with the H-AP in one group and not separated into two groups. Thus, the SI at the H-AP does not occur, and this leads to the UL SNR for MU$_{k}$ being defined as
\begin{equation}
\gamma^{UL}_{k,\hat{l}}=\frac{P^{UL}_{k,\hat{l}}\vert\mathbf{v}^{H}_{k,\hat{l}}\mathbf{\hat{g}}_{k,\hat{l}}\vert^2}{\sum_{j\in \mathcal{S}_l,j\neq k}P^{UL}_{j,\hat{l}}\vert\mathbf{v}^{H}_{k,\hat{l}}\mathbf{\hat{g}}_{j,\hat{l}}\vert^2+\sigma^{2}_{\hat{l}}\Vert\mathbf{v}^{H}_{k,\hat{l}}\Vert^{2}}.
\end{equation} 
In the case of HD-WPCN, some of the solutions derived for the FD-WPCN in the previous section can be reused. This include $\mathbf{w}^{\star}_{k,l}$, $\vartheta^{UL\star}_{k,\hat{l}}$, $P^{UL\star}_{k,\hat{l}}$, and $\tau^{\star}_l$. However, the solution for $\mathbf{v}^{\star}_{k,l}$ should be modified to
\begin{equation}
\label{equTbeamWPCN1}
\mathbf{v}^{\star}_{k,l}=\Big(\sum_{j\in \mathcal{S}_{\hat{l}}}P^{UL}_{j,l}\mathbf{\hat{g}}_{j,l}\mathbf{\hat{g}}^{H}_{j,l}+ \sigma^{2}_{l}\mathbf{I}_{M} \Big)^{-1}\sqrt{P^{UL}_{k,l}}\mathbf{\hat{g}}_{k,l}, 
\end{equation} 
due to the absence of RSI. 
\begin{figure}[t!]
\centering
\includegraphics[width=3.2in]{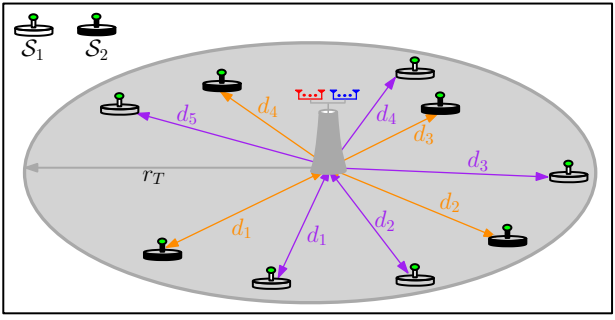}
\caption{Schematic representation of node distribution used for simulation.}
\label{figplotdiag}
\end{figure}
\subsection{FD-WPCN and HD-WPCN implementation comparison}
\label{subsecimpcom}
In this subsection, some insightful observations are made by comparing the FD-WPCN to the HD-WPCN. First, as the number of MUs increases, the FD-WPCN will have better UL SNR for each MU. This is because, the $\sum_{j\in \mathcal{S}_l,j\neq k}P^{UL}_{j,\hat{l}}\vert\mathbf{v}^{H}_{k,\hat{l}}\mathbf{\hat{g}}_{j,\hat{l}}\vert^2$ component of the noise for the SNR consists of approximately $K-1$ and $K/2-1$ interfering signals from other MUs operating in UL communication for the HD-WPCN and FD-WPCN, respectively. Also, from equation (\ref{equTbeamWPCN}), it can be observed that as the number of MUs increases, the RSI reduces for the FD-WPCN system. Having a higher $\sum_{j\in \mathcal{S}_l,j\neq k}P^{UL}_{j,\hat{l}}\vert\mathbf{v}^{H}_{k,\hat{l}}\mathbf{\hat{g}}_{j,\hat{l}}\vert^2$ for the HD-WPCN and reducing RSI in the FD-WPCN, the FD-WPCN has a better SNR compared to the HD-WPCN.

Next, the effect of increasing the H-AP transmit power for both the HD-WPCN and the FD-WPCN is discussed. For the case of HD-WPCN, the sum-rate will be saturated at the high source power regime due to the presence of the channel error and non-linear EH behavior. This happens because as the H-AP power is increased, the amount of harvested energy at each MU increases until it saturates. With an increase in harvested energy at each node, there is an increase in the transmit power of each MU. The increase in MU transmit power means the $\sum_{j\in \mathcal{S}_l,j\neq k}P^{UL}_{j,\hat{l}}\vert\mathbf{v}^{H}_{k,\hat{l}}\mathbf{\hat{g}}_{j,\hat{l}}\vert^2$ component of the MU UL SNR in (\ref{equSNRWPCN}) increases. Hence, the UL SNR saturates for the high transmit power range due to the noise component becoming more significant than the actual information signal. For the FD-WPCN system, the sum-rate saturates and degrades compared to the HD-WPCN system. This is because the FD-WPCN is influenced by the channel error, RSI, and the non-linear EH nature of the MUs.
\section{Simulation Results and Discussion}
\label{secresults}
\begin{figure}[t!]
\centering
\includegraphics[width=3.4in]{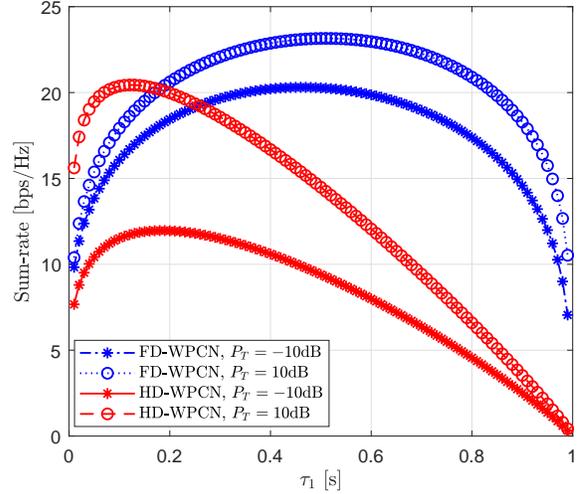}
\caption{A plot of sum-rate against varying $\tau_{1}$  over one random channel generation (i.e. $P^{DL}_{0,max}=-10$ dB and $10$ dB, $r_{T}=10$ m, $K=4$ and $M=4$).}
\label{figtimeperform}
\end{figure}
In this section, simulation results for the proposed sum-rate optimization algorithms are presented. The MUs are assumed to be randomly and uniformly distributed around the H-AP within a circle of radius $r_{T}$, as shown in Fig. \ref{figplotdiag}. $\sigma^{2}_{E}$ for the simulations is set to be either $0$, $0.01$, or $0.1$ for each plot. The EH efficiency of the MUs is assumed to be $70\%$ ( i.e., $\beta_k=0.7$). If not stated, the loop-back noise power and the H-AP receiver antenna noise power are assumed to be $-80$ dBm, and each MU antenna saturation threshold power is set at $7$ dBm. The attenuation constant is set to $C_{0}=-10$ dB. A comparison of the FD-WPCN algorithm with that of HD-WPCN is presented throughout this section. Also, a comparison of the FD-WPCN and HD-WPCN optimal schemes to their suboptimal fixed $\tau_l$ schemes are considered in the simulations\footnote{There are no comparison with existing work because, during the submission of this work, existing models considered TDMA and not TDD mode of operation.}. For the suboptimal fixed scheme, $\tau_1=\tau_2=0.5$ is assumed. The results presented in this section were acquired from running $10^{3}$ random channel realizations. 
\begin{figure}[t!]
\centering
\includegraphics[width=3.4in]{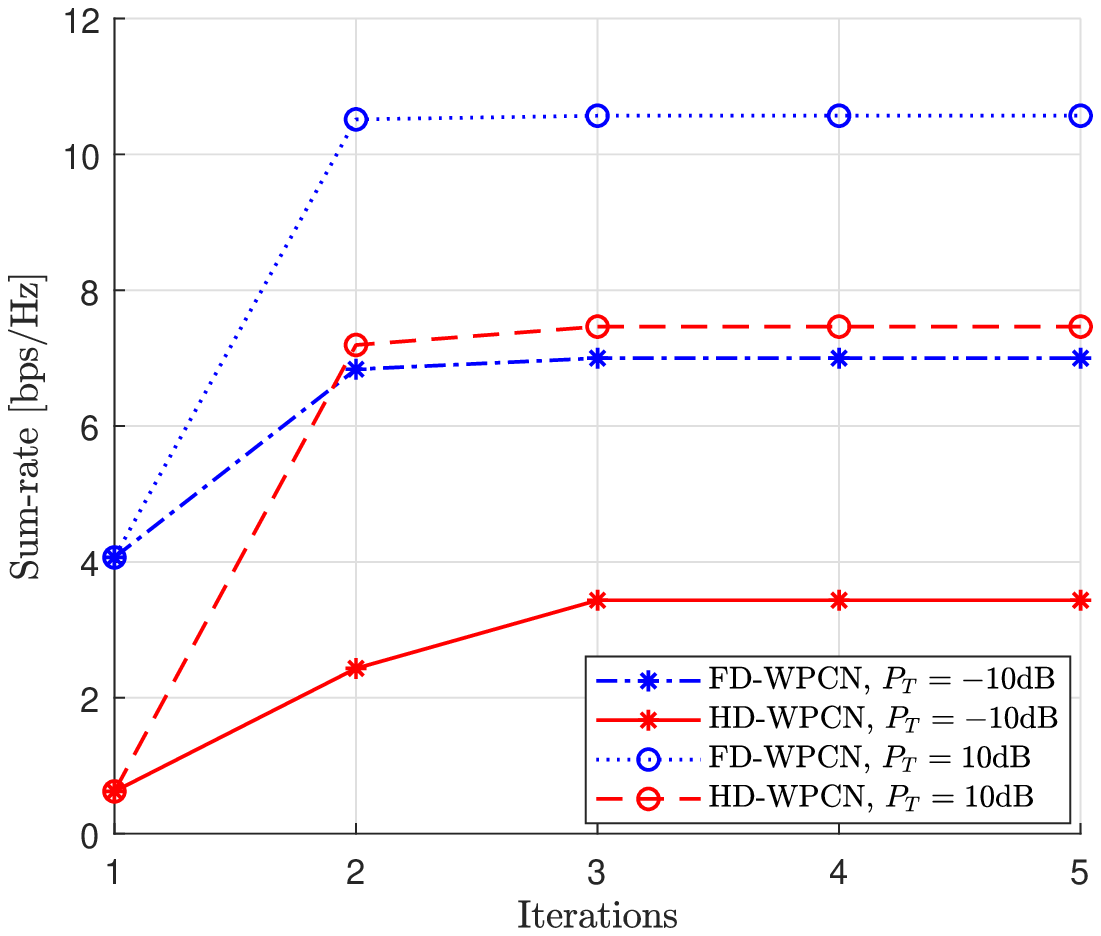}
\caption{A plot of sum-rate against number of Algorithm \ref{algMGSAlgorithmWPCN} convergence iterations over one random channel generation (i.e. $P^{DL}_{0,max}=-10$ dB and $10$ dB, $\tau_{1}=0.5$ s, $r_{T}=10$ m, $K=4$ and $M=4$).}
\label{figmmseperform}
\end{figure}
\begin{figure}[t!]
\centering
\includegraphics[width=3.4in]{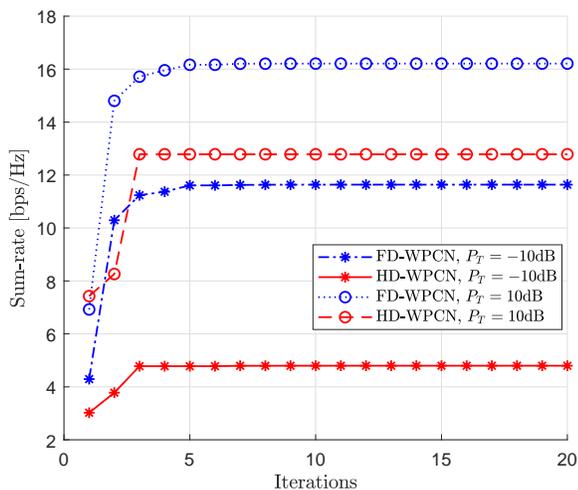}
\caption{A plot of sum-rate against number of Algorithm \ref{algMGSAlgorithmWPCN1} iterations over one random channel generation (i.e. $P^{DL}_{0,max}=-10$ dB and $10$ dB, $\tau_{1}=0.5$ s, $r_{T}=10$ m, $K=4$ and $M=4$).}
\label{figrandperform}
\end{figure}
\subsection{Algorithm behavior}
\label{subsecalgoperform}
In this subsection, simulation results are presented in order to show the behavior of Algorithms \ref{algMGSAlgorithmWPCN} and \ref{algMGSAlgorithmWPCN1} in terms of convexity and convergence. Fig. \ref{figtimeperform} is a graphical proof of problem (\ref{equoptmain2WPCN}) being a convex problem of $\tau_{l}$ for H-AP source powers of $-10$ dB and $10$ dB. Therefore, the global optimality of Algorithm \ref{algMGSAlgorithmWPCN1} can be found using a simple line search method. Figs. \ref{figmmseperform} and \ref{figrandperform} show the convergence plots for Algorithms \ref{algMGSAlgorithmWPCN} and \ref{algMGSAlgorithmWPCN1}, respectively. From Fig. \ref{figmmseperform}, the sum-rate (i.e., Algorithm \ref{algMGSAlgorithmWPCN} - WMMSE approach) converges very fast with a few number of iterations (less than $5$). This implies that the sum-rate is maximized with optimal $a^{\star}_{k,l}$, $P^{\star}_{k,l}$, and $\mathbf{w}^{\star}_{k,l}$ with just a few iterative runs of Algorithm \ref{algMGSAlgorithmWPCN}. Finally, Fig. \ref{figrandperform} shows the convergence of Algorithm \ref{algMGSAlgorithmWPCN1}. In this paper, the line search method selected for Algorithm \ref{algMGSAlgorithmWPCN1} is the GLSM. From Fig. \ref{algMGSAlgorithmWPCN1}, it is observed that the GLSM converges at the global optimum sum-rate after less than $10$ iterations. Please, note that if another line search method (e.g., gradient decent method) is used, the number of iterations needed for convergence may either reduce or increase. For either FD-WPCN or HD-WPCN, the same computational complexity is needed to implement both algorithms. However, less computation is need for Algorithm \ref{algMGSAlgorithmWPCN} compared to Algorithm \ref{algMGSAlgorithmWPCN1}. For example, Algorithm \ref{algMGSAlgorithmWPCN} uses $3$ iterations to achieve a suboptimal rate. However, Algorithm \ref{algMGSAlgorithmWPCN1} uses about $24$ iterations (i.e., Algorithm \ref{algMGSAlgorithmWPCN} iterations $\times$ Algorithm \ref{algMGSAlgorithmWPCN1} iterations $=3\times 8=24$) to find the optimal sum-rate. Hence, in practical systems, an engineer can decide which algorithm to implement based on the performance requirements of deployed nodes, such as, computational power, QoS and resource management.
\subsection{EH discussion}
\label{subsecenergyharvestsimeresults}
\begin{figure}[t!]
\centering
\includegraphics[width=3.4in]{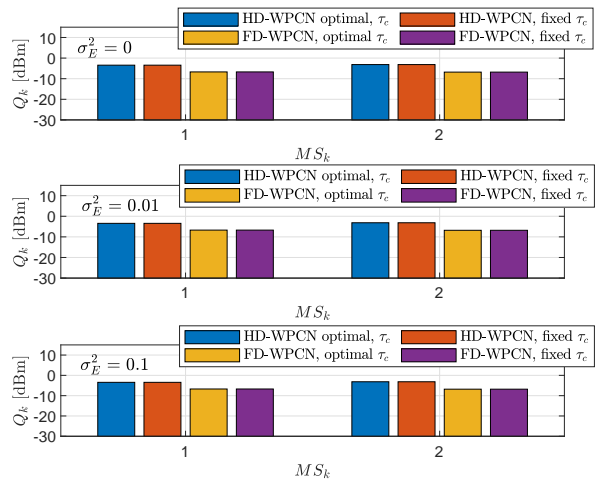}
\caption{A plot of harvested energy at each MU, $MU_{k}$, where $P^{DL}_{0,max}=0$ dB, $r_{T}=10$ m, $K=2$ and $M=2$.}
\label{figEHpernode}
\end{figure}
Simulation results and discussions on the amount of harvested energy at each node are presented in this subsection. In this subsection, $K=2$ MUs are used for the simulation results. The effect of the CSI estimation error variance on the amount of harvested energy at each node is investigated with Fig. \ref{figEHpernode}. With an H-AP transmit power of $0$ dB and CSI estimation error variances of $[0,0.01,0.1]$, each MU$_{k}$ harvests the same amount of power for the different error variances. This is because the amount of harvested energy at each node is not influenced by the estimated channel, but rather the actual channel. Therefore, the EH circuit at each node does not require the CSI to undergo EH. From Fig. \ref{figEHpernode}, considering either the HD-WPCN or the FD-WPCN mode, the amount of harvested energy at each node is the same for all the MUs. This is due to the beamforming vector at the H-AP used for each MU during the EH phase. In addition to this observation, the optimal and suboptimal schemes for both WPCN modes harvest the same amount of energy. Finally, the HD-WPCN mode requires more energy to achieve maximum sum-rate compared to the FD-WPCN mode. This system performance can be explained by the number of nodes present in a group. The HD-WPCN has an extra MU node present in the EH phase. Hence, MU$_{1}$ and MU$_{2}$ take advantage of each other's beamformed RF signals. However, this is not so with the FD-WPCN mode, since MU$_{1}$ and MU$_{2}$ operate either WET or WIT in a phase. Hence, each MU can not take advantage of the beamformed RF signal to the other MU. 
\begin{figure}[t!]
\centering
\includegraphics[width=3.4in]{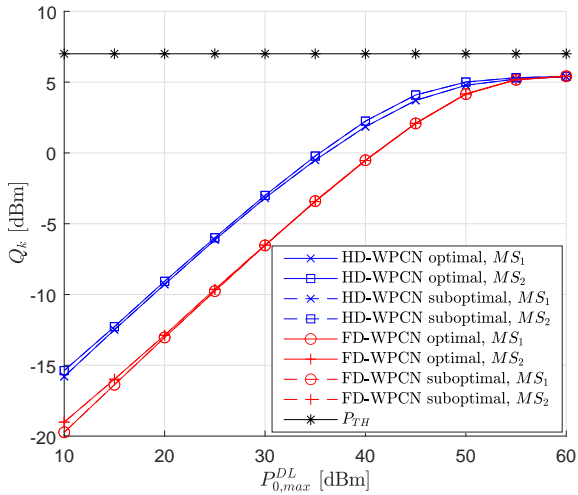}
\caption{A plot of harvested energy at each $MU_{k}$  against increasing H-AP power, $P^{DL}_{0,max}$, with $r_{T}=10$ m, $K=2$ and $M=2$.}
\label{figEHwithpower}
\end{figure}

\begin{figure}[t!]
\centering
\includegraphics[width=3.4in]{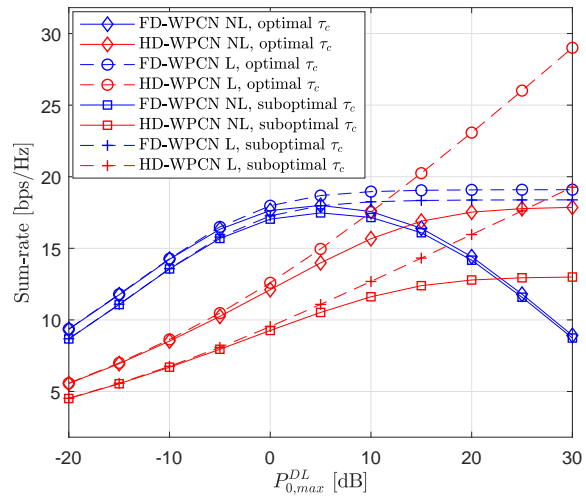}
\caption{A plot of sum-rate against H-AP transmit power, $P^{DL}_{0,max}$ where $\tau_{1}=0.5$, $r_{T}=10$m, $K=4$ and $M=4$.}
\label{figlinearperform}
\end{figure}

\begin{figure}[t!]
\centering
\includegraphics[width=3.4in]{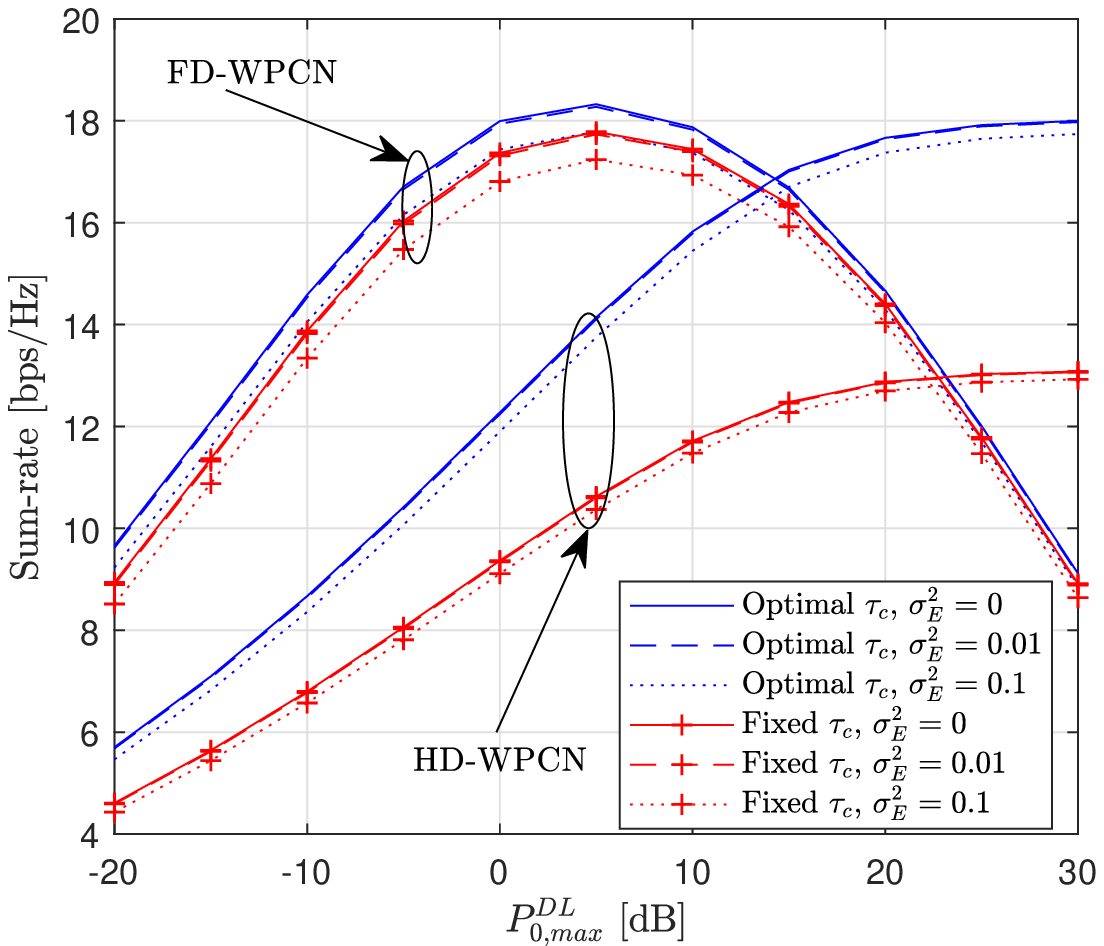}
\caption{A plot of sum-rate against H-AP transmit power, $P^{DL}_{0,max}$ with $r_{T}=10$m, $K=4$ and $M=4$.}
\label{figsumrate1}
\end{figure}

The illustration of the influence of the non-linear EH model on each MU node's harvested energy is presented in Fig. \ref{figEHwithpower}. From Fig. \ref{figEHwithpower}, as the H-AP transmit power increases, the amount of harvested energy at each MU increases. However, the amount of harvested energy saturates when the amount of energy received at the MU antenna is above its antenna saturation power value. It can be observed from Fig. \ref{figEHwithpower} that the amount of harvested energy reaches a constant value of about $5$ dBm at a H-AP transmit power of $50$ dBm and $55$ dBm for the HD-WPCN and FD-WPCN methods, respectively. This implies that, at the antenna power saturation point, each MU will harvest the same amount of energy for both the FD-WPCN and HD-WPCN transmission modes. The behavior of the linear and non-linear EH model are presented and discussed using Fig. \ref{figlinearperform}. For the HD-WPCN mode, it can be observed that for the linear model, the sum-rate increases with increasing H-AP transmit power. However, the non-linear EH model attains sum-rate saturation. This behavior is due to the harvested energy being constant at the antenna saturation power point at each MU. Hence, the UL power for each MU also becomes constant, as seen in Fig. \ref{figEHwithpower}. This constant MU transmit power affects the UL sum-rate negatively when it comes to the FD-WPCN system. Because the H-AP source power keeps increasing, the SI also increases, while the MU UL power remains constant. Hence, the sum-rate achieved by the FD-WPCN scheme will begin to reduce at a certain H-AP transmit power. It is, therefore, prudent to use H-AP transmit power that will not cause redundancy in the UL sum-rate, as seen in Figs. \ref{figlinearperform} and \ref{figsumrate1}\footnote{The sum-rate is a convex problem with respect to $P^{DL}_{0,max}$, assuming $\sum_{j\in \mathcal{S}_{\hat{l}}}\Vert\mathbf{w}_{j,\hat{l}}\Vert^2=P^{DL}_{0,max}$ in (\ref{equSNRWPCN}) is made constant. However, a closed-form solution for this problem is hard to solve. Hence, to determine the maximum sum-rate point based on $P^{DL}_{0,max}$, a line search method can be used.}. However, in the FD-WPCN, it is preferable for the H-AP to transmit a level of power that can achieve a sum-rate equivalent to the sum-rate peak point. But, by increasing the H-AP transmit power, the sum-rate of the FD-WPCN linear model attains saturation due to RSI. Therefore, the use of the non-linear EH model mimics real-world/practical system scenarios.
\subsection{System achievable sum-rate discussion}
\label{subsecsum-ratesimeresults}
\begin{figure}[t!]
\centering
\includegraphics[width=3.4in]{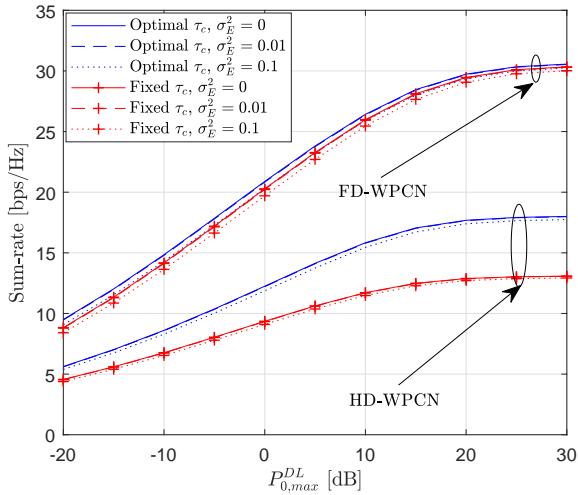}
\caption{A plot of sum-rate against H-AP transmit power with no RSI, $P^{DL}_{0,max}$ with $r_{T}=10$m, $K=4$ and $M=4$.}
\label{figsumrate2}
\end{figure}

\begin{figure}[t!]
\centering
\includegraphics[width=3.4in]{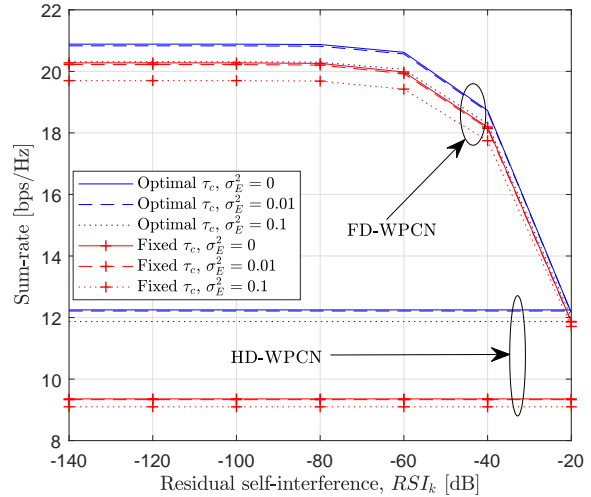}
\caption{A plot of sum-rate against RSI with $\hat{P}^{DL}_{0,max}=0$ dB, $r_{T}=10$m, $K=4$, fixed $\tau_l =0.5$, and $M=4$.}
\label{figRSIrate}
\end{figure}

\begin{figure}[t!]
\centering
\includegraphics[width=3.4in]{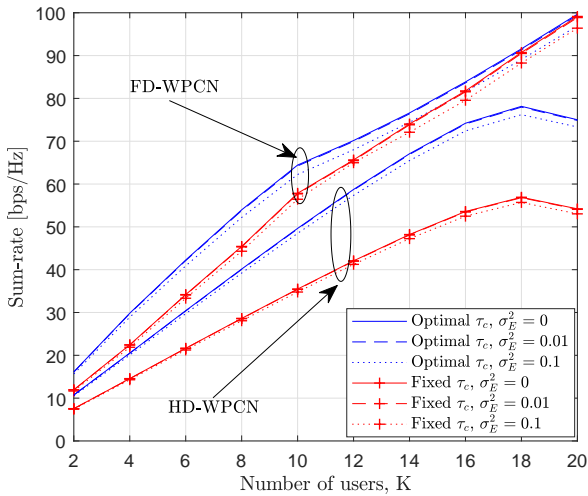}
\caption{A plot of sum-rate against increasing number of users with $\hat{P}^{DL}_{0,max}=0$ dB, $r_{T}=10$m, $K=4$, $M=4$, and fixed $\tau_l =0.5$).}
\label{figNumrate}
\end{figure}

In this subsection, simulation results for the maximum system achievable sum-rate are presented. In Fig. \ref{figsumrate1}, it is observed that the performance of the FD-WPCN exceeds that of the HD-WPCN at the $P^{DL}_{0,max}<15$ dB region. However, in the $P^{DL}_{0,max}>15$ dB region, both HD-WPCN schemes outperform their FD-WPCN counterparts. This is because in the $P^{DL}_{0,max}>15$ dB region, the performance FD-WPCN schemes are affected by RSI, MU UL power, and channel error effect. The FD-WPCN achieves a reduction in sum-rate for  $P^{DL}_{0,max}>15$ dB for all values of $\sigma^{2}_{E}$ used in our simulations. HD-WPCN schemes, however, attain constant sum-rate from  $P^{DL}_{0,max}>25$ dB. Also, as expected, the optimal schemes for both FD-WPCN and HD-WPCN perform better compared to their suboptimal counterparts. This implies that one may need more advanced SIC and CSI estimation techniques to implement the proposed FD and HD WPCN schemes, as seen in Fig. \ref{figsumrate2}\footnote{Note that with Fig. \ref{figsumrate2}, perfect SIC is assumed.}.
\begin{figure}[t!]
\centering
\includegraphics[width=3.4in]{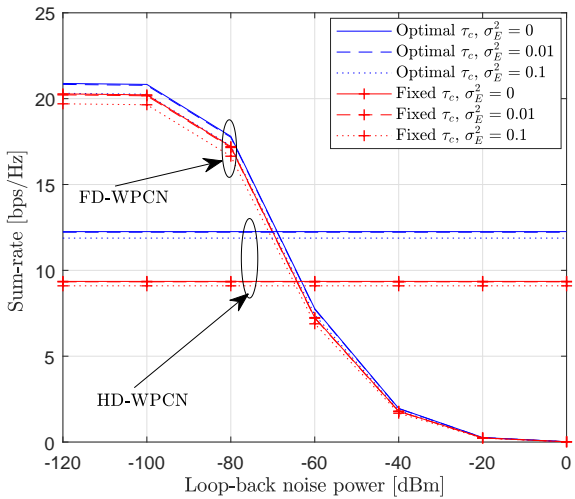}
\caption{A plot of sum-rate against loop-back channel noise power,  $\hat{\sigma}^{2}_{\hat{H}}$, where $\hat{P}^{DL}_{0,max}=0$ dB, $r_{T}=10$m, $K=4$, fixed $\tau_l =0.5$, and $M=4$.}
\label{figVarrate}
\end{figure}
Figs. \ref{figVarrate} and \ref{figRSIrate} show average sum-rate against the loop-back noise power and RSI, respectively. From both plots, it is observed that as the overall RSI increases, the system achievable sum-rate reduces for both the optimal and suboptimal FD-WPCN schemes. This is due to the increase in SI occurring at the H-AP. For both figures, the HD-WPCN schemes have a constant sum-rate because it is not affected by the loop-back noise power and in turn the RSI. For the loop-back noise power, the decrease in sum-rate is experienced at noise values greater than $-100$ dBm, as shown in Fig. \ref{figVarrate}. Similarly, the decrease in the sum-rate begins at a RSI of $-80$ dB in Fig. \ref{figRSIrate}. From Fig. \ref{figVarrate}, the FD-WPCN schemes (i.e., both optimal and suboptimal) outperform the HD-WPCN schemes at loop-back noise power less than $-70$ dBm. This emphasizes the need for a better SIC or a hybrid FD-WPCN and HD-WPCN system for better system performance. Finally, there are apparent differences between the performance gain of the perfect CSI case as compared to the imperfect CSI case. The gain differences are due to the compounded effect of channel estimation errors. To avoid this issue, the channel estimation technique implemented must reduce the estimation error drastically, which is a well-known fact in channel estimation research works \cite{Yoo06,Xiang12}. Fig. \ref{figNumrate} shows the sum-rate improvement with an increasing number of MUs, $K$, for both HD-WPCN and FD-WPCN. With an increasing number of MUs and a fixed transmit SNR, the FD-WPCN is a better technique for communication compared to the HD-WPCN even with the presence of RSI. However, with better SIC techniques, the FD-WPCN schemes will perform better. The FD-WPCN schemes approach similar performance when $K>16$, but the HD-WPCN schemes see a reduction in performance at $K>18$.
\section{Conclusion}
\label{secconclusion}
Resource allocation for a multi-user WPCN network in which MUs communicate with FD AP is investigated in this paper. The system model considered two groups of MUs based on their current communication mode with the FD H-AP and channel access. The optimization of the time resource allocation, channel allocation, transmit and receive beamformers are implemented to maximize the UL sum-rate. It is shown that with the proposed iterative algorithm the optimal sum-rate can be achieved. The simulation results presented covered the comparison of the proposed FD-WPCN with a HD-WPCN system model. The FD-WPCN outperformed the HD-WPCN system model at low transmit SNR in terms of average sum-rate. HD-WPCN has a dominant performance at high transmit SNR in terms of average sum-rate due to RSI at the FD-WPCN H-AP. 

The work presented in this paper considered both perfect and imperfect channel estimation, MIMO H-AP configuration and WPCN EH. The perfect channel estimation approach gave better sum-rate performance over its imperfect counterpart. In addition, the smaller the channel estimation error, the better the system performance. Hence, a promising future extension can consider better channel estimation and SIC techniques to improve the system performance. To improve the sum-rate and the amount of harvested energy at each node, massive MIMO H-AP and MIMO MUs can be considered as future extensions of this work. Other WPT techniques such as SWIPT TS and PS schemes can be researched in place of the WPCN method considered in this work. With the SWIPT techniques, information signals can be transmitted in both phases of communication to the AP and MUs. This implies that the MUs will not only transmit information but can receive information, hence improving sum-rate and system performance. 
\appendices
\section{Proof of Theorem 1}\label{AppApp1}
The stepwise approach used in acquiring the optimal energy beamformer for the H-AP is presented here. The approach used in \cite{Hoon16} and \cite{Hoon15} is adopted to attain optimal solution in this paper. The differential of the Lagrangian with respect to $\mathbf{w}_{k,l}$ and KKT conditions from (\ref{equoptmainWPCN}) are
\begin{equation}
\label{equApp1}
\mathbf{w}_{k,l}\bigg[\frac{\tau_{l} }{\tau_{\hat{l}}} \sum_{j\in \mathcal{S}_{l}}\lambda^{UL}_{j}\beta_{j}\Big[\mathbf{h}_{j,l}\mathbf{h}_{j,l}^{H}\Big]-\lambda^{DL}\mathbf{I}_{M}\bigg] =0,
\end{equation} 
\begin{equation}
\label{equApp2}
\lambda^{DL}\bigg(P^{DL}_{0,max}- \sum^{K}_{k=1}\sum^{2}_{l=1}\Vert\mathbf{w}_{k,l}\Vert^{2}\bigg) =0,
\end{equation}
\begin{equation}
\label{equApp3}
\sum^{K}_{k=1}\lambda^{UL}_{k}\beta_{k}\sum^{2}_{l=1}\bigg(\frac{\tau_{l} }{\tau_{\hat{l}}}\sum_{j\in \mathcal{S}_{l}}\Big[\vert\mathbf{h}_{k,l}^{H}\mathbf{w}_{k,l}\vert^{2}\Big]-P^{UL}_{k,l} \bigg) =0,
\end{equation}
and
\begin{equation}
\label{equApp4}
\lambda^{UL}_{k}\geq 0, \lambda^{DL}\geq 0.
\end{equation}
Denote $\mathbf{A}=\frac{\tau_{l} }{\tau_{\hat{l}}} \sum_{j\in \mathcal{S}_{l}}\lambda^{UL}_{j}\beta_{j}\Big[\mathbf{h}_{j,l}\mathbf{h}_{j,l}^{H}\Big]-\lambda^{DL} \mathbf{I}_{M}$ and $\mathbf{B}=\frac{\tau_{l} }{\tau_{\hat{l}}} \sum_{j\in \mathcal{S}_{l}}\lambda^{UL}_{j}\beta_{j} \Big[\mathbf{h}_{j,l}\mathbf{h}_{j,l}^{H}\Big]$. If $\lambda^{DL} > 0$ and $\lambda^{UL}_{k}=0$, $\mathbf{w}_{k,l}=0$ from (\ref{equApp1}) since $\mathbf{A}=-\lambda^{DL} \mathbf{I}_{M}$. This solution is infeasible because it contradicts  the complementary slackness condition (\ref{equApp2}) where $\lambda^{DL}$ must not be zero. Also, for the case where $\lambda^{DL}=0$ and $\lambda^{UL}_{k}>0$, $\mathbf{w}_{k,l}=0$ from (\ref{equApp1}) which contradicts the complementary slackness condition (\ref{equApp3}). Hence both $\lambda^{DL}$ and $\lambda^{UL}_{k}$ are greater than zero. 

Let the eigenvalue decomposition of matrix $\mathbf{A}$ be $\mathbf{A}=\mathbf{U}_{B}\left( \Lambda_{B}-\lambda^{DL}\mathbf{I}_{M}\right) \mathbf{U}^{H}_{B}$, with eigenvector matrix and eigenvalue matrix $\mathbf{U}_{B}\in \mathbb{C}^{MxM}$, and $\Lambda_{B}=$diag$(\lambda_{B,1},\ldots,\lambda_{B,K})$ with $\lambda_{B,1}\geq\ldots\geq\lambda_{B,K}$, respectively. $\mathbf{B}$ is always positive semi-definite because $\lambda_{k}^{UL}>0$ with non-negative eigenvalues (i.e. $\lambda_{B,1}\geq\ldots\geq\lambda_{B,K}>0$). For $\mathbf{A}$ to have non-positive eigenvalues (i.e. $\lambda_{B,1}-\lambda^{DL} \leq 0$), $\lambda^{DL}\geq\lambda_{B,1}>0$. 

If $\lambda^{DL}>\lambda_{B,1}$, $\mathbf{w}_{k,l}=0$, and $\mathbf{A}$ is full-rank and negative-definite contradicting (\ref{equApp2}), hence $\lambda^{DL*}=\lambda_{B,1}$. From this deduction, the null space of $\mathbf{A}$ spans the unit-norm eigenvector $\mathbf{u}_{B,1}$ of $\mathbf{B}$ with the largest eigenvalue $\lambda_{B,1}$, (i.e. $\mathbf{A}\mathbf{u}_{B,1}=0$). Hence, $\mathbf{w}_{k,l}=\Omega\mathbf{u}_{B,1}$ is justified for attaining the optimal conditions. $\Omega=\sqrt{P^{DL}_{0,max}}$ which is acquired from the complementary slackness condition (\ref{equApp2}) with $\lambda^{DL}>0$.  Note that in this work, imperfect channel is assumed at the H-AP, that is, only the estimated channel (i.e., $\hat{h}_{j,l}$) is known at the H-AP. Hence, $\mathbf{u}_{B,1}$ is determined using $\hat{h}_{j,l}$ instead of $h_{j,l}=\hat{h}_{j,l}+\tilde{h}_{j,l}$. However, for the assumption of perfect channel estimation, $h_{j,l}=\hat{h}_{j,l}+\tilde{h}_{j,l}$ is used to determine $\mathbf{u}_{B,1}$. In addition, the optimal $\mathbf{w}^{\star}_{k,c}$ derived here is used for the power saturation of the EH case. $\blacksquare$
\section{Proof of Convexity for WMMSE Problem}\label{AppApp2a}
The Lagrangian and KKT conditions for the problem (\ref{equWMMSEWPCN}) are deduced as
\begin{equation}
\begin{aligned}
&\mathcal{L}(\mathbf{v}_{k,l},\vartheta^{UL}_{k,l},P^{UL}_{k,l},\lambda^{UL}_{k})\\&\text{ }
\begin{aligned}
&&&&=\sum^{K}_{k=1}\sum^{2}_{l=1}a_{k,\hat{l}}\Big(\vartheta^{UL}_{k,l}\Big(1-\sqrt{P^{UL}_{k,l}}\Big(\mathbf{v}^{H}_{k,l}\mathbf{\hat{g}}_{k,l}+\mathbf{\hat{g}}^{H}_{k,l}\mathbf{v}_{k,l}\Big)
\end{aligned}
\\&\text{ }
\begin{aligned}
&&&&&&&&+\sum_{j\in \mathcal{S}_{\hat{l}}}P^{UL}_{j,l}\vert\mathbf{v}^{H}_{k,l}\mathbf{\hat{g}}_{j,l}\vert^{2}+\Big( \sum_{j\in \mathcal{S}_{l}}\Vert\mathbf{w}_{j,l}\Vert^{2}\hat{\sigma}^{2}_{\hat{H}}+\sigma^{2}_{l}\Big)\times
\end{aligned}
\\&\text{ }
\begin{aligned}
&&&&&&&&&&&&&\Vert\mathbf{v}_{k,l}\Vert^{2}\Big)-\log \vartheta^{UL}_{k,l}\Big)
\end{aligned}
\\&\text{ }
\begin{aligned}
&&&+\sum^{K}_{k=1}\lambda^{UL}_{k}\beta_{k}\sum^{2}_{l=1}\bigg(\frac{\tau_{l} }{\tau_{\hat{l}}}\sum_{j\in \mathcal{S}_{l}}\Big[\vert\mathbf{h}_{k,l}^{H}\mathbf{w}_{k,l}\vert^{2}\Big]-P^{UL}_{k,l} \bigg),
\end{aligned}
\end{aligned}
\end{equation}
\begin{equation}
\begin{aligned}
&\frac{\partial\mathcal{L}}{\partial\vartheta^{UL}_{k,l}}=a_{k,\hat{l}}-\Big(a_{k,\hat{l}}\sqrt{P^{UL}_{k,l}}\Big(\mathbf{v}^{H}_{k,l}\mathbf{\hat{g}}_{k,l}+\mathbf{\hat{g}}^{H}_{k,l}\mathbf{v}_{k,l}\Big)+\frac{a_{k,\hat{l}}}{\vartheta^{UL}_{k,l}}\Big)
\\&\text{ }
\begin{aligned}
&&&&&&&&&&& +a_{k,\hat{l}}\Big( \sum_{j\in \mathcal{S}_{l}}\Vert\mathbf{w}_{j,l}\Vert^{2}\hat{\sigma}^{2}_{\hat{H}}+\sigma^{2}_{l}\Big)\Vert\mathbf{v}_{k,l}\Vert^{2}
\end{aligned}
\\&\text{ }
\begin{aligned}
&&&&&&&&&&&&&&+a_{k,\hat{l}}\sum_{j\in \mathcal{S}_{\hat{l}}}P^{UL}_{j,l}\vert\mathbf{v}^{H}_{k,l}\mathbf{\hat{g}}_{j,l}\vert^{2},
\end{aligned}
\end{aligned}
\end{equation}
\begin{equation}
\begin{aligned}
&\!\frac{\partial\mathcal{L}}{\partial P^{UL}_{k,l}}=a_{k,\hat{l}}\vartheta^{UL}_{k,l}\Big(\sum_{j\in \mathcal{S}_{\hat{l}}}\vert\mathbf{v}^{H}_{j,l}\mathbf{\hat{g}}_{k,l}\vert^{2} -\frac{\mathbf{v}^{H}_{k,l}\mathbf{\hat{g}}_{k,l}}{\sqrt{P^{UL}_{k,l}}}\Big)-\lambda^{UL}_{k},
\end{aligned}
\end{equation}
\begin{equation}
\begin{aligned}
&\frac{\partial\mathcal{L}}{\partial \mathbf{v}_{k,l}}=2a_{k,\hat{l}}\vartheta^{UL}_{k,l}\Big( \sum_{j\in \mathcal{S}_{l}}\Vert\mathbf{w}_{j,l}\Vert^{2}\hat{\sigma}^{2}_{\hat{H}}+\sigma^{2}_{l}\Big)\mathbf{v}_{k,l}
\\&\text{ }
\begin{aligned}
&&&&&&&&&&&+2a_{k,\hat{l}}\vartheta^{UL}_{k,l}\mathbf{v}_{k,l}\sum_{j\in \mathcal{S}_{\hat{l}}}\sqrt{P^{UL}_{k,l}}\mathbf{\hat{g}}_{k,l}\mathbf{\hat{g}}^{H}_{k,l}
\end{aligned}
\\&\text{ }
\begin{aligned}
&&&&&&&&&&&&&&&& -2a_{k,\hat{l}}\vartheta^{UL}_{k,l}\sqrt{P^{UL}_{k,l}}\mathbf{\hat{g}}_{k,l}.
\end{aligned}
\end{aligned}
\end{equation}
Next, the second derivatives for problem (\ref{equWMMSEWPCN}) are shown as follows
\begin{equation}
\begin{aligned}
&\quad\frac{\partial^{2}\mathcal{L}}{\partial\vartheta^{UL^{2}}_{k,l}}=\frac{a_{k,\hat{l}}}{\vartheta^{UL^2}_{k,l}},\text{ }\text{ }\text{ }\text{ }\text{ }\text{ }\frac{\partial^{2}\mathcal{L}}{\partial P^{UL^2}_{k,l}}=\frac{a_{k,\hat{l}}\vartheta^{UL}_{k,l}}{P^{UL}_{k,l}}\mathbf{v}^{H}_{k,l}\mathbf{\hat{g}}_{k,l},
\end{aligned}
\end{equation}
\begin{equation}
\begin{aligned}
&\frac{\partial^{2}\mathcal{L}}{\partial \mathbf{v}^{2}_{k,l}}=2a_{k,\hat{l}}\vartheta^{UL}_{k,l}\Big( \sum_{j\in \mathcal{S}_{l}}\Vert\mathbf{w}_{j,l}\Vert^{2}\hat{\sigma}^{2}_{\hat{H}}+\sigma^{2}_{l}\Big)\mathbf{I}\\&\text{ }
\begin{aligned}
&&&&&&&&&&+2a_{k,\hat{l}}\vartheta^{UL}_{k,l}\sum_{j\in \mathcal{S}_{\hat{l}}}\sqrt{P^{UL}_{k,l}}\mathbf{\hat{g}}_{k,l}\mathbf{\hat{g}}^{H}_{k,l},
\end{aligned}
\end{aligned}
\end{equation}
\begin{equation}
\begin{aligned}
&\quad\frac{\partial^{2}\mathcal{L}}{\partial\vartheta^{UL}_{k,l}\partial P^{UL}_{k,l}}=\frac{ a_{k,\hat{l}}\mathbf{v}^{H}_{k,l}\mathbf{\hat{g}}_{k,l}}{\sqrt{P^{UL}_{k,l}}}+a_{k,\hat{l}}\sum_{j\in \mathcal{S}_{\hat{l}}}\vert\mathbf{v}^{H}_{k,l}\mathbf{\hat{g}}_{j,l}\vert^{2},
\end{aligned}
\end{equation}
\begin{equation}
\begin{aligned}
&\!\frac{\partial^{2}\mathcal{L}}{\partial P^{UL}_{k,l} \partial \mathbf{v}_{k,l} }=a_{k,\hat{l}}\vartheta^{UL}_{k,l}\mathbf{\hat{g}}_{k,l}\Big( \frac{2\mathbf{\hat{g}}^{H}_{k,l}\mathbf{v}_{j,l}\sqrt{P^{UL}_{k,l}}-1}{\sqrt{P^{UL}_{k,l}}}\Big),
\end{aligned}
\end{equation}
\begin{equation}
\begin{aligned}
&\frac{\partial\mathcal{L}}{\partial \mathbf{v}_{k,l}\partial\vartheta^{UL}_{k,l}}=2a_{k,\hat{l}}\Big( \sum_{j\in \mathcal{S}_{l}}\Vert\mathbf{w}_{j,l}\Vert^{2}\hat{\sigma}^{2}_{\hat{H}}+\sigma^{2}_{l}\Big)\mathbf{v}_{k,l}\\&\text{ }
\begin{aligned}
&&&&&&&&&&&&&&+2a_{k,\hat{l}}\mathbf{v}_{k,l}\sum_{j\in \mathcal{S}_{\hat{l}}}\sqrt{P^{UL}_{k,l}}\mathbf{\hat{g}}_{k,l}\mathbf{\hat{g}}^{H}_{k,l}
\end{aligned}
\\&\text{ }
\begin{aligned}
&&&&&&&&&&&&&&&&&&-2a_{k,\hat{l}}\vartheta^{UL}_{k,l}\sqrt{P^{UL}_{k,l}}\mathbf{\hat{g}}_{k,l}.
\end{aligned}
\end{aligned}
\end{equation}
From the above solutions, it can be observed that all the second derivatives are positive. Hence, the presented problem is concave with respect to all the variable. Therefore, the proposed algorithm converges to a global optimum solution. $\blacksquare$
\section{Proof of Lemma 1}\label{AppApp2}
The Lagrangian of problem (\ref{equoptmain2WPCN}) after simplification is given as 
\begin{equation}
\begin{aligned}
&\mathcal{L}(\tau_{l},\lambda^{UL}_{k})=\sum^{K}_{k=1}\sum^{2}_{l=1}a_{k,\hat{l}}\tau_{l}\log_2(1+\gamma^{UL}_{k,l})+\varphi\bigg(1-\sum^{2}_{l=1}\tau_{l}\bigg)\\&\text{ }
\begin{aligned}
&&&&&&&&&&&+\sum^{K}_{k=1}\sum^{2}_{l=1}\frac{\lambda^{UL}_{k}\beta_{k}\tau_{l}}{1-\tau_{l}} \sum_{j\in \mathcal{S}_{l}}\vert\mathbf{h}_{k,l}^{H} \mathbf{w}_{j,l}\vert^{2}
 \end{aligned}\\&\text{ }
\begin{aligned}
&&&&&&&&&&&-\sum^{K}_{k=1}\sum^{2}_{l=1}\lambda^{UL}_{k}P^{UL}_{k,l}. 
\end{aligned}
\end{aligned}
\end{equation}
The first differential of the Lagrangian is deduced as
\begin{equation}
\begin{aligned}
&\frac{\partial\mathcal{L}}{\partial\tau_{l}}=\sum^{K}_{k=1}a_{k,\hat{l}}\log_2(1+\gamma^{UL}_{k,l})+\varphi\\&\text{ }
\begin{aligned}
&&&&&&&&&-\frac{1}{(1-\tau_{l})^2} \sum^{K}_{k=1}\lambda^{UL}_{k}\beta_{k} \sum_{j\in \mathcal{S}_{l}}\vert\mathbf{h}_{k,l}^{H} \mathbf{w}_{j,l}\vert^{2}. 
\end{aligned}
\end{aligned}
\end{equation}
To confirm that the Lagrangian is a convex function, the second derivative is found to be 
\begin{equation}
\begin{aligned}
\frac{\partial^2\mathcal{L}}{\partial^2\tau_{l}}= - \frac{2}{(1-\tau_{l})^3} \sum^{K}_{k=1}\lambda^{UL}_{k}\beta_{k}\sum_{j\in \mathcal{S}_{l}}\vert\mathbf{h}_{k,l}^{H} \mathbf{w}_{j,l}\vert^{2}. 
\end{aligned}
\end{equation}
Since the second derivative of Lagrangian with respect to $\tau_l$ has a negative value, the Lagrangian is a concave function. This holds at MU antenna saturation power too, with the second derivative given as
\begin{equation}
\begin{aligned}
\frac{\partial^2\mathcal{L}}{\partial^2\tau_{l}}= - \frac{2}{(1-\tau_{l})^3} \sum^{K}_{k=1}\lambda^{UL}_{k}\beta_{k}P_{TH}.
\end{aligned}
\end{equation}
$\blacksquare$
\ifCLASSOPTIONcaptionsoff
  \newpage
\fi
\bibliographystyle{IEEEtr}

\end{document}